\documentclass{nuws}
\usepackage{amsmath,amssymb,url}
 \newif\ifpdf 
 \ifx\pdfoutput\undefined
 \pdffalse    
 \else
 \pdfoutput=1 
 \pdftrue
 \fi \ifpdf   
     \usepackage[pdftex]{graphicx} 
 \else
     \usepackage{graphicx}         
 \fi

\newcommand{\Red}{}
\newcommand{\Black}{}

\def\w{4.38cm} \def\gw{8.82cm} \def\W{8.51cm} 
\urldef{\FLUKA-3D}\url{http://www.mi.infn.it/~battist/neutrino.html}
\urldef{\Workshop}\url{http://www-zeuthen.desy.de/nuastro/
                   publications/sources/workshop2001/icpost_tr.html}

\begin{document}

\title{Atmospheric muons and neutrinos}
\author[1]{Vadim A. Naumov}
\affil[1]{Dipartimento di Fisica and Sezione INFN di Ferrara,
          Via del Paradiso 12, I-44100 Ferrara, Italy and \\
          Laboratory for Theoretical Physics, Irkutsk State
          University, Gagarin boulevard 20, RU-664003 Irkutsk,
          Russia}
\correspondence{V.\,A.\,Naumov (naumov@fe.infn.it)}

\firstpage{1}
\pubyear{2001}


\maketitle

\begin{abstract}
This paper is a mini-review of the atmospheric muon and neutrino
flux calculations based upon a recent version of {CORT} code and
up-to-date data on primary cosmic rays and hadronic interactions.
A comparison of calculations with a representative set of
atmospheric muon data for momenta below $\sim1$ TeV/c is presented.
The overall agreement between the calculated muon fluxes and the
data provides an evidence in favor of the validity of adopted
description of hadronic interactions and shower development.
In particular, this supports the low-energy atmospheric neutrino
fluxes predicted with CORT which are essentially lower than those
used in current analyses of the sub-GeV and multi-GeV neutrino
induced events in underground neutrino detectors.
\end{abstract}

\protect\section{Introduction}
\label{sec:Introduction}

In recent paper by \citet{Fiorentini01a} an updated version of
FORTRAN code {CORT} ({\bf C}osmic-{\bf O}rigin {\bf R}adiati\-on
{\bf T}ransport) has been described and some results of low and
intermediate energy muon and neutrino flux calculations performed
with {CORT} have been discussed. A detailed comparison with the muon
fluxes and charge ratios measured in several modern balloon-borne
experiments suggests that the atmospheric neutrino flux is
essentially lower than one used in the current analyses of the
sub-GeV and multi-GeV neutrino induced events in underground
detectors. Some additional results which confirm this conclusion were
discussed in \citep{Fiorentini01b}.

Present work can be considered as an {\em addendum} to the
above-mentioned papers. Its main goal is to provide a comparison of
the predicted muon fluxes with a more representative set of data from
balloon-borne and ground-based experiments, including rather old (but
by no means outdated), which cover extensive ranges of muon momenta,%
\footnote{In this short write-up we limit ourselves with a selected
          set of muon data at momenta $p\lesssim1$ TeV/c (some of
          these data were taken from the recent compilations by
          \citet{Vulpescu98,Vulpescu01} and \citet{Hebbeker01}) and
          do not consider the underground and underwater muon data.
          Additional data can be found in the comprehensive
          reference manual by \citet{Grieder01}, and also in Refs.
          \citep{Bugaev98,Naumov98a,Naumov00a,%
          Sinegovskaya01,Naumov01b}.}
zenith angles and atmospheric depths. Such a comparison may be useful
to obtain further insight into the atmospheric neutrino problem.

\protect\section{The CORT code}
\label{sec:CORT}

Like the earlier versions of {CORT}
\citep{Naumov84,Bugaev84,Bugaev85,Bugaev86,Bugaev87a,Bugaev87b,%
Bugaev88,Bugaev89a,Bugaev90,Naumov93,Bugaev98},
the new Fortran 90 code implements a numerical integration of a
system of one-dimensional kinetic equations describing the production
and transport of nuclei, nucleons, light mesons, muons, neutrinos,
and antineutrinos of low and intermediate energies in the atmosphere.
It takes into account solar modulation, geomagnetic and
meteorological effects, energy loss of charged particles, muon
polarization and depolarization effects. The exact relativistic
kinematics is applied for description of particle interactions and
decays.  The new code has a number of options and run modes. It is
rather flexible and permits fast modification of many input data with
intrinsic switch keys.

In order to evaluate geomagnetic effects and to take into account the
anisotropy of the primary cosmic-ray flux in the vicinity of the
Earth, CORT uses the method of Ref. \citep{Naumov84} and detailed
maps of the effective vertical cutoff rigidities by \citet{Dorman71}.
The maps are corrected for the geomagnetic pole drift and compared
with the later results reviewed by \citet{Smart94} and with the
recent AMS 
data the on the proton flux in near Earth orbit \citep{Alcaraz00a}.
The interpolation between the reference points of the maps is
performed by means of two-dimensional local B-spline. The Quenby-Wenk
relation \citep{Dorman71}, re-normalized to the vertical cutoffs, is
applied for evaluating the effective cutoffs for oblique directions.
More sophisticated effects, like the short-period variations of the
geomagnetic field, Forbush decrease, re-entrant cosmic-ray albedo
contribution, etc., are neglected. We also neglect the geomagnetic
bending of the trajectories of charged secondaries and multiple
scattering effects. Validity of our treatment of propagation of
secondary nucleons and nuclei has been verified using all available
data on secondary proton and neutron spectra in the atmosphere
\citep{Naumov84,Bugaev84,Bugaev85,Naumov00b,Naumov01a}.

The meteorological effects are included using the Dorman model of the
atmosphere \citep{Dorman72} which assumes an isothermal stratosphere
and constant gradient of temperature (as a function of depths) below
the tropopause. Ionization, radiative and photonuclear muon energy
losses are treated as continuous processes. This approximation is
quite tolerable for atmospheric depths $h\lesssim2\times10^3$
g/cm${}^2$ at all energies of interest \citep{Naumov94}. Propagation
of $\mu^+$ and $\mu^-$ originating from every source (pion or kaon
decay) is described by separate kinetic equations for muons with
definite polarization at production. These equations automatically
account for muon depolarization through the energy loss (but not
through the Coulomb scattering).

\protect\section{Primary cosmic ray spectrum and composition}
\label{sec:Primaries}

In the present calculations, the nuclear component of primary cosmic
rays is broken up into 5 principal groups: H, He, CNO, Ne-S and Fe
with average atomic masses $A$ of 1, 4, 15, 27 and 56, respectively.
We do not take into account the isotopic composition of the primary
nuclei and assume $Z=A/2$ for $A>1$, since the expected effect on the
secondary lepton fluxes is estimated to be small with respect to
present-day experimental uncertainties in the absolute cosmic-ray
flux and chemical composition.

We parametrize the spectra of the H and He groups at $E<120$
GeV/nucleon by fitting the data of the balloon-borne experiment BESS
obtained by a flight in 1998 \citep{Sanuki00,Sanuki01a}.
For higher energies (but below the knee) we use the data by a series
of twelve balloon flights of JACEE \citep{Asakimori98} and the result
of an analysis by \citet{Wiebel-Sooth98} based upon a representative
compilation of world data on primaries. Since the fits obtained by
\citet{Wiebel-Sooth98} turned out to be very close to the combined
result of the JACEE experiments, the model is called ``BESS+JACEE''
fit.

\begin{figure}[h]
\centering
\ifpdf
\includegraphics[width=\W]{ps_pr.pdf} 
\includegraphics[width=\W]{ps_he.pdf} 
\else
\includegraphics[width=\W]{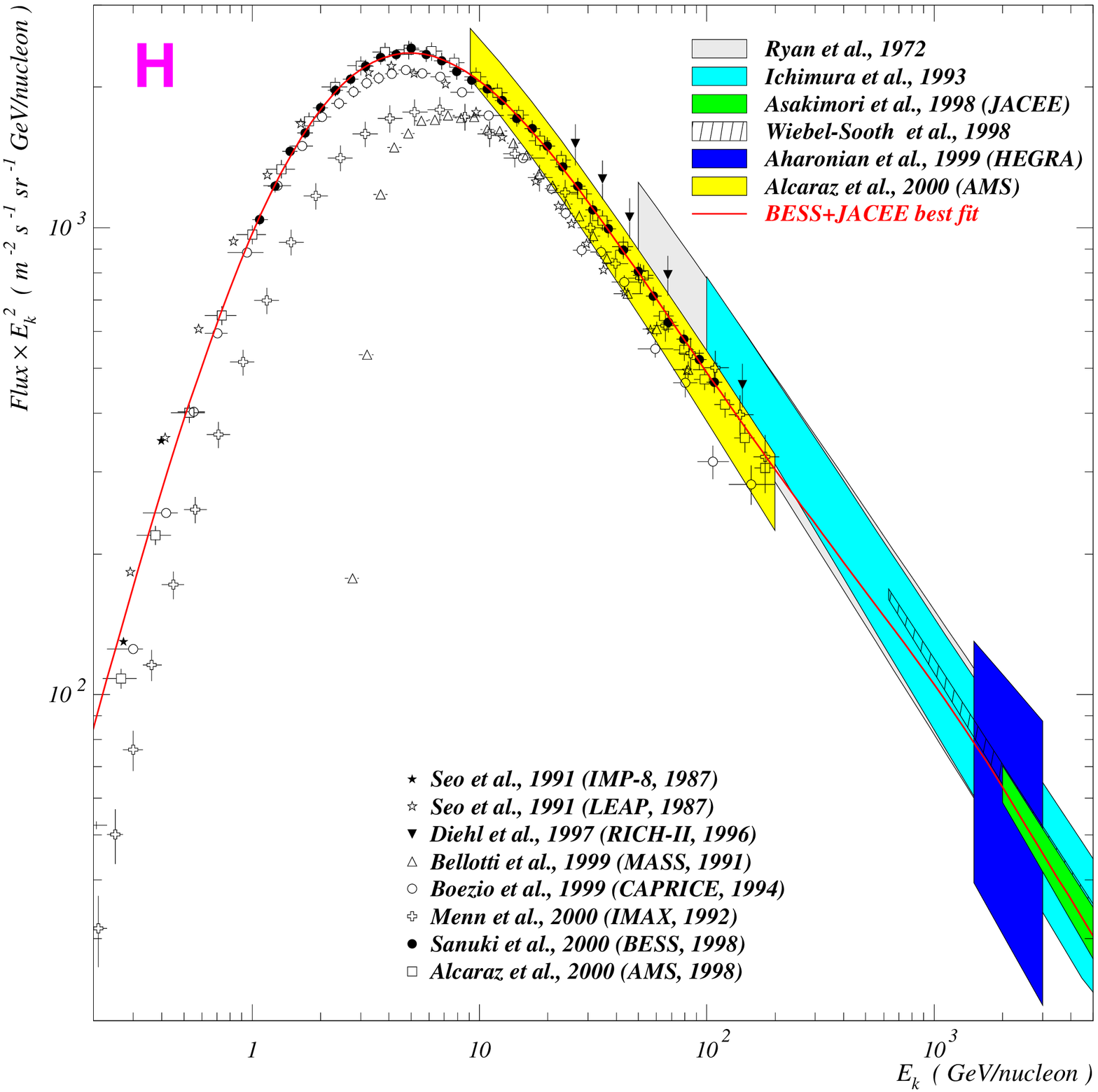} 
\includegraphics[width=\W]{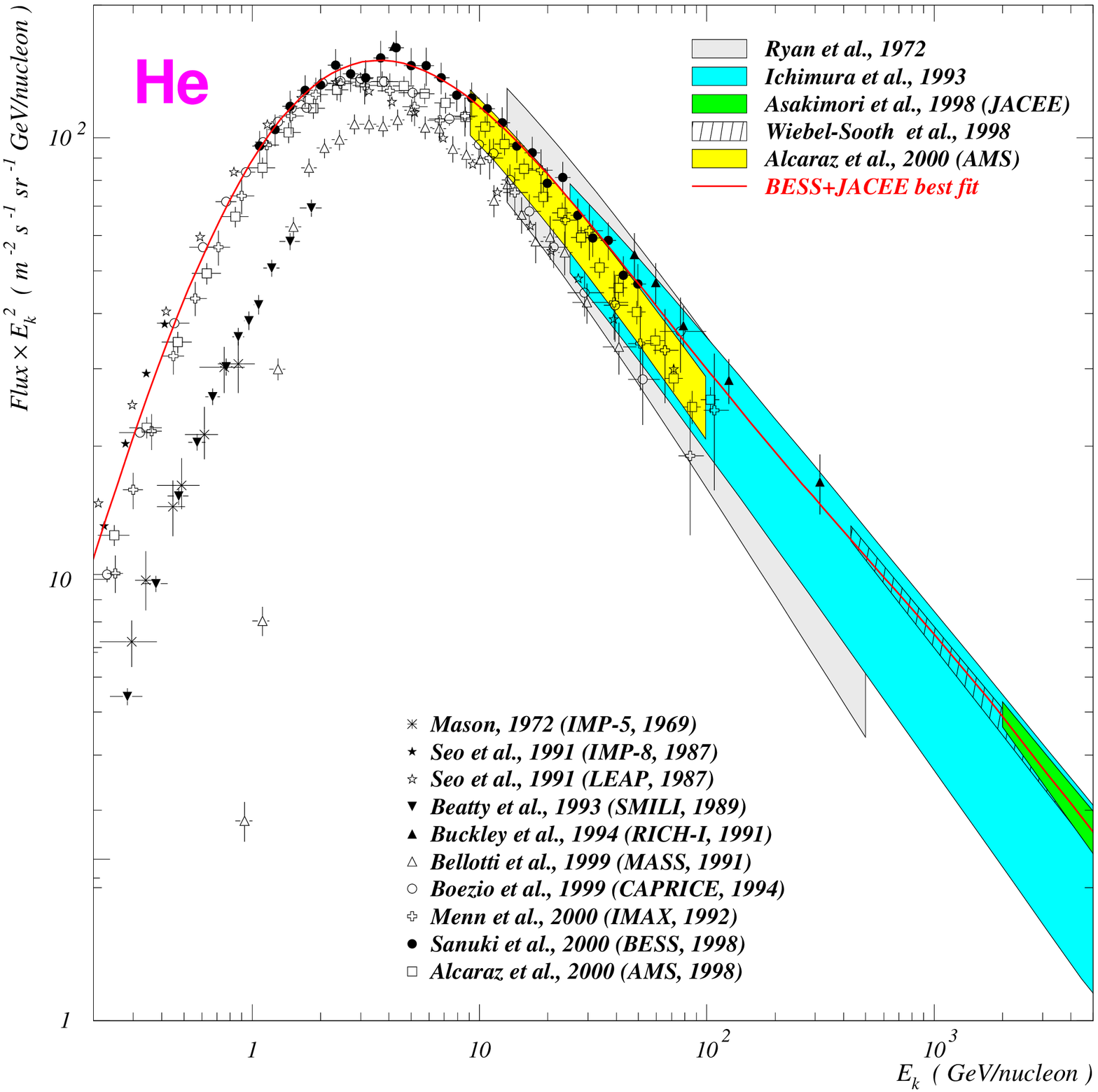} 
\fi
\protect\caption{Differential spectra of protons (top panel) and
                 helium nuclei (bottom panel) as a function of
                 kinetic energy per nucleon.
                 The data points and filled/shaded areas are for
                 the experimental data and power-law fits of the
                 data as is shown in the legends.
                 The solid curves represent the BESS+JACEE best fit
                 from \citet{Fiorentini01a} (see text for details).
\label{f:PS}}
\end{figure}

Figure \ref{f:PS} shows a comparison between the BESS+JACEE fit,
the data from \citet{Sanuki00,Sanuki01a} (BESS) and
\citet{Asakimori98} (JACEE), the fit from \citet{Wiebel-Sooth98}
(shaded areas) and the data from several other experiments
\citep{Mason72,Ryan72,Seo91,Beatty93,Ichimura93,Buckley94,Diehl97,%
      Menn00,Aharonian99,Bellotti99,Boezio99a,Alcaraz00b,Alcaraz00c},
performed in different periods of solar activity. The filled
areas in Fig. \ref{f:PS} represent power-type parametrizations
of the spectra derived by
\citet{Asakimori98,Ryan72,Ichimura93,Aharonian99,Alcaraz00b,%
      Alcaraz00c}
from the original data. The legends indicate the publication dates
and (when relevant) the years of measurements.

We assume that the spectra of the remaining three nuclear groups are
similar to the helium spectrum. This assumption does not contradict
the world data for the CNO and Ne-S nuclear groups but works a bit
worse for the iron group.
Nevertheless, a more sophisticated model would be unpractical since
the corresponding correction would affect the secondary lepton fluxes
by a negligible margin.

In this paper we do not consider the effects of solar modulation.
Therefore the predicted muon and neutrino fluxes are to some extent
the maximum ones possible within our approach.

\protect\section{Nucleon-nucleus and nucleus-nucleus interactions}
\label{sec:interactions}

All calculations with the earlier version of CORT
were based on semiempirical models for inclusive nucleon and meson
production in collisions of nucleons with nuclei by
\citet{Kimel'74,Kimel'75,Serov73}.%
\footnote{See also Refs. \citet{Kalinovsky85,Sychev99} for the most
          recent versions.}
The Kimel'-Mokhov (KM) model is valid for projectile nucleon momenta
above $\sim4$ GeV/c and for the secondary nucleon, pion and kaon
momenta above 450, 150 and 300 MeV/c, respectively. Outside these
ranges (that is mainly within the region of resonance production of
pions) the Serov-Sychev (SS) model was used.

Both models are in essence comprehensive parametrizations of the
relevant accelerator data. It is believed that the combined
``KM+SS'' model provides a safe and model-inde\-pendent basis to the
low-energy atmospheric muon and neutrino calculations.
However it is not free of uncertainties.
For the present calculation, the fitting
parameters of the KM model for meson and nucleon production off
different nuclear targets were updated using accelerator data not
available for the original analysis
\citep{Kimel'74,Kimel'75,Kalinovsky85}. The values of the parameters
were extrapolated to the air nuclei (N, O, Ar, C).  The overall
correction is less than 10-15\% within the kinematic regions
significant to atmospheric cascade calculations. Besides that
energy-dependent correction factors were introduced into the model to
tune up the output $\pi^+/\pi^-$ ratio taking into account the
relevant new data.

The processes of meson regeneration and charge exchange
($\pi^\pm+\mathrm{Air}\to\pi^{\pm(\mp)}+X$ etc.) are not of critical
importance for production of leptons with energies of our interest
and can be considered in a simplified way. Here we use a proper
renormalization of the meson interaction lengths, which was deduced
from the results by \citet{Vall86} obtained for high-energy cascades.

The next important ingredient of any cascade calculations is a model
for nucleus-nucleus collisions. Here we consider a modest
generalization of a simple ``Glauber-like'' model used in
\citep{Naumov84,Bugaev85}. Namely, we write the inclusive spectrum of
secondary particles $c$ ($c=p$, $n$, $\pi^\pm$, $K^\pm$,
$K^0,\ldots$) produced in collisions of nuclei as
\begin{align*}
\frac{dN_{\mathrm{AB} \to cX}}{d x} = & \xi_{\mathrm{AB}}^c
\left[Z \frac{dN_{p\mathrm{B} \to cX}}{dx}+
   (A-Z)\frac{d N_{n\mathrm{B} \to cX}}{dx}\right]\nonumber\\
&+\left(1-\xi_{\mathrm{AB}}^c\right)
\left[Z\delta_{cp}+(A-Z)\delta_{cn}\right]\delta(1-x).
\end{align*}
Here $dN_{N\mathrm{B} \to cX}/dx$ is the spectrum of particles $c$
produced in a collision of a nucleon $N$ ($N=p,n$) with nucleus
B, $x$ is the Feynman variable, and $\xi_{\mathrm{AB}}^c$ is the
average fraction of inelastically interacting nucleons of
the projectile nucleus A. The term proportional to delta function
describes the contribution of ``spectator'' nucleons from the
projectile nucleus.

In the standard Glauber-Gribov multiple scattering theory the
quantity $\xi_{\mathrm{AB}}^c$ is certainly independent of the type
of inclusive particle $c$. On the other hand, it depends of the type
of nucleus collision. Indeed, essentially all nucleons participate in
the central AB collisions ($\xi_{\mathrm{AB}}^c\simeq1$)%
\footnote{Here we assume for simplicity that the atomic weight of
          the projectile nucleus is not much larger than that of
          the target nucleus.}
while, according to the well-known Bialas--Bleszy\'nski--Czy\.z (BBC)
relation \citep{Bialas76},
$\xi_{\mathrm{AB}}^c=\sigma_{N\mathrm{B}}^{\mathrm{inel}}/
                     \sigma_{\mathrm{AB}}^{\mathrm{inel}}$
for the minimum bias collisions.

To use the above model in a cascade calculation one should take into
account that nucleons and mesons are effectively produced in nuclear
collisions of different kind. Namely, the contribution from central
collisions is almost inessential for the nucleon component of the
cascade but quite important for light meson production. Thus one can
expect that effectively
$\xi_{\mathrm{AB}}^{\pi,K}>\xi_{\mathrm{AB}}^{p,n}$. We use the BBC
relation for nucleon production by any nucleus while for meson
production we put $\xi_{\mathrm{He-Air}}^{\pi,K}=\xi$, where $\xi$ is
a free parameter. Variations of this parameter within the
experimental limits yield a comparatively small effect to the muon
fluxes (except for very high altitudes) and inessential
($\lesssim6$\%) effect to the neutrino fluxes at sea level
\citep{Fiorentini01a}. Effect of similar variations of the parameters
$\xi_{\mathrm{A-Air}}^{\pi,K}$ for other nuclear groups is completely
negligible.

\protect\section{Numerical results for muons}
\label{sec:NumRes}

\begin{figure*}[t]
\centering
\ifpdf
\includegraphics[width=\gw]{c94f.pdf}~
\includegraphics[width=\gw]{c98f.pdf} 
\else
\includegraphics[width=\gw]{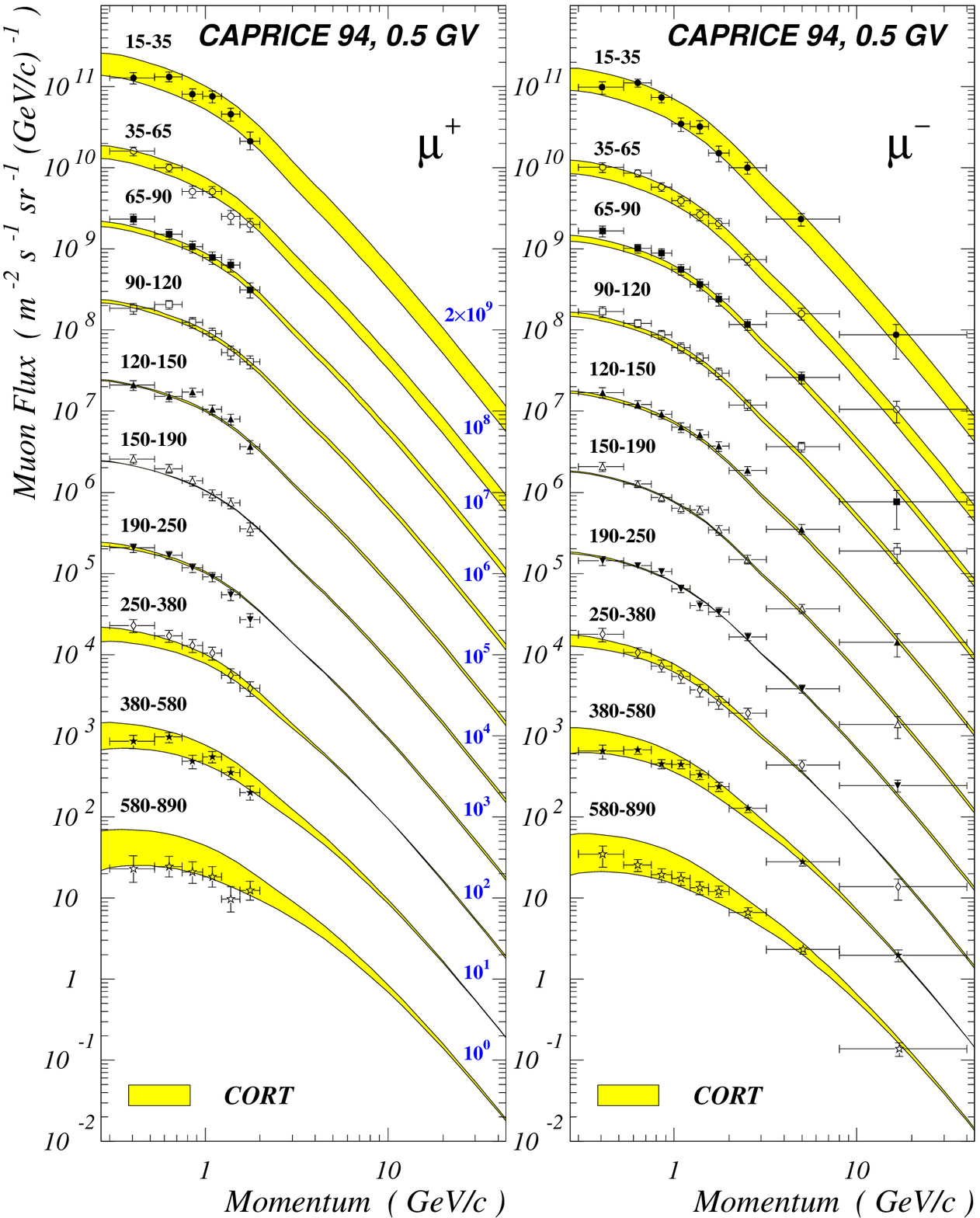}~
\includegraphics[width=\gw]{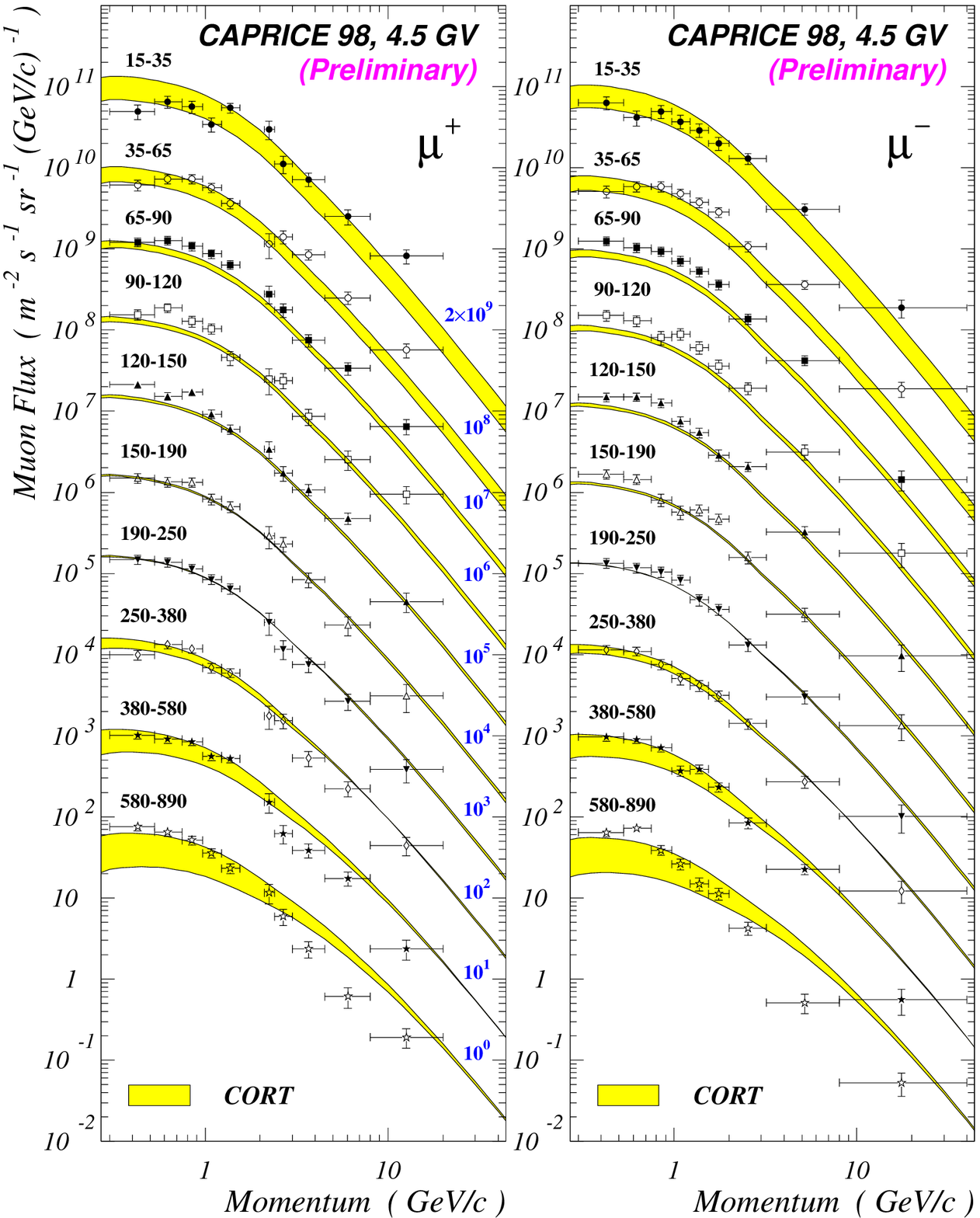} 
\fi
\protect\caption{Differential momentum spectra of $\mu^+$ and $\mu^-$
                 for 10 atmospheric depth ranges $\Delta h_i$
                 (are indicated on the left of each panel).
                 The data points are from the CAPRICE\,94 experiment
                 \citep{Boezio99b,Boezio00} and from the
                 {\em preliminary} analysis of the CAPRICE\,98
                 experiment \citep{Hansen01}.
                 The filled areas display the expected variations
                 of the muon fluxes within the ranges $\Delta h_i$.
                 All the data are scaled with the factors
                 indicated at the first and third panels on the
                 right.
\label{f:CAPRICE94-98}}
\end{figure*}

Figure \ref{f:CAPRICE94-98} displays a comparison of the
calculated momentum spectra of $\mu^+$ and $\mu^-$ for 10 atmospheric
depth ranges $\Delta h_i= 15-35, 35-65, 65-90, 90-120, 120-150,
150-190, 190-250, 250-380, 380-580$, and $580-890$ g/cm${}^2$ with
the data of two balloon-borne experiments CAPRICE\,94
\citep{Boezio99b,Boezio00} and CAPRICE\,98 \citep{Hansen01}
(the latter experimental data are {\em preliminary}).
The nominal geomagnetic cutoff rigidity $R_c$ is about 0.5 GV
(4.5 GV) and the detection cone is about $20^\circ$ ($14^\circ$)
around the vertical direction with the average incident angle of
about $9^\circ$ ($8^\circ$) for CAPRICE\,94 (CAPRICE\,98).
The data points in Fig. \ref{f:CAPRICE94-98} are the muon
intensities at the Flux-weighted Average Depths (FAD) defined in
\citep{Boezio99b}, while the filled areas display the calculated
variations of the muon spectra inside the ranges $\Delta h_i$
($i=1,\ldots,10$). Namely, they are obtained by considering the
minimal and maximal muon fluxes within each range $\Delta h_i$.
The calculations are performed for the conditions of the
experiment. The parameter $\xi$ is taken to be 0.685
(the best value, according to \citet{Fiorentini01a}).%
\footnote{It has been shown \citep{Fiorentini01a} that
          the indetermination of $\xi$ is only
          significant for $h<(15-20)$ g/cm${}^2$.}

\clearpage 
\begin{figure*}[b]
\centering
\ifpdf
\includegraphics[width=\linewidth]{cgl.pdf} 
\else
\includegraphics[width=\linewidth]{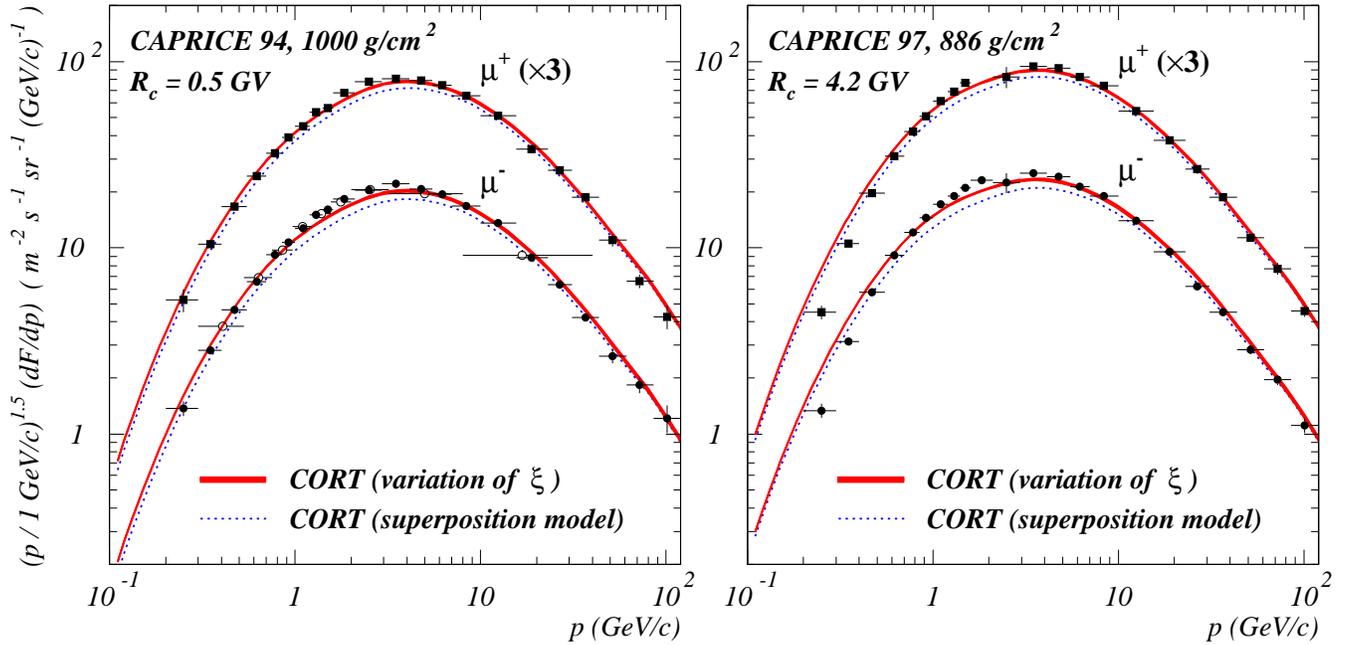} 
\fi
\protect\caption{Differential momentum spectra of positive and
                 negative muons at the ground level.
                 The data points are from the experiments
                 CAPRICE\,94 \citep{Boezio99b,Boezio00,Kremer99} and
                 CAPRICE\,97 \citep{Kremer99}. The $\mu^+$ data are
                 scaled with a factor of 3. The solid (narrow) bands
                 and dotted curves are calculated with CORT as is
                 explained in legends.
\label{f:CAPRICE94-97}}
\end{figure*}
\clearpage 

It is important to note that the thickness of the bands is relatively
small just for the region of effective muon and neutrino production
that is in the neighborhood of the broad maxima of the muon flux
($100-300$~g/cm${}^2$). By this is meant that, in this region,
an error in evaluation of the FAD cannot introduce essential
uncertainty. Outside the region of effective production of leptons,
the amplitude of the muon flux variations increases with decreasing
muon momenta on account for the strong dependence of the meson
production rate upon the depth and the growing role of the muon
energy loss and decay at $h\gtrsim300$~g/cm${}^2$.

Figure \ref{f:CAPRICE94-97} shows a comparison of the calculated
differential momentum spectra of $\mu^+$ and $\mu^-$ with the
most accurate current data obtained at ground level in the
experiments CAPRICE\,94 \citep{Boezio99b,Boezio00,Kremer99}
($h=1000$ g/cm$^2$, $R_c=0.5$ GV) and CAPRICE\,97 \citep{Kremer99}
($h=886$ g/cm$^2$, $R_c=4.2$ GV).
The detection cone in both experiments was the same as described
above for CAPRICE\,94. The calculation are done by using the KM+SS
interaction model. Variation of the parameter $\xi$ from 0.517
(the BBC value) to 0.710 (an experimental upper limit derived from
the data on interactions of $\alpha$ particles with light nuclei).%
\footnote{See, e.g., Ref. \citep{Kowalski81} and references
         therein.}
leads to the almost negligible ($\lesssim3$\%) effect in the
ground-level muon flux.
For reference, the spectra calculated by using the superposition
model for nucleus-nucleus interactions are also added. The data
below $\sim10$ GeV/c seem to be precise enough in order to
conclude that they disfavor the superposition model.

As is seen from Figs. \ref{f:CAPRICE94-98} and \ref{f:CAPRICE94-97},
there is a substantial agreement between the calculations with CORT
and the data of the CAPRICE experiments within a wide range
atmospheric depths. This agreement is {\em none the worse} than it
is for the recent 3D Monte Carlo calculations \citep{Battistoni01a,%
Engel01,Honda01b,Liu01,Poirier01,Wentz01a,Wentz01b}.%
\footnote{See also Ref. \citep{Fiorentini01b} for the direct
          comparison with the FLUKA 3D result.}

\begin{figure}[htb]
\centering
\ifpdf
\includegraphics[width=8.4cm]{geof.pdf} 
\includegraphics[width=8.4cm]{geor.pdf} 
\else
\includegraphics[width=8.4cm]{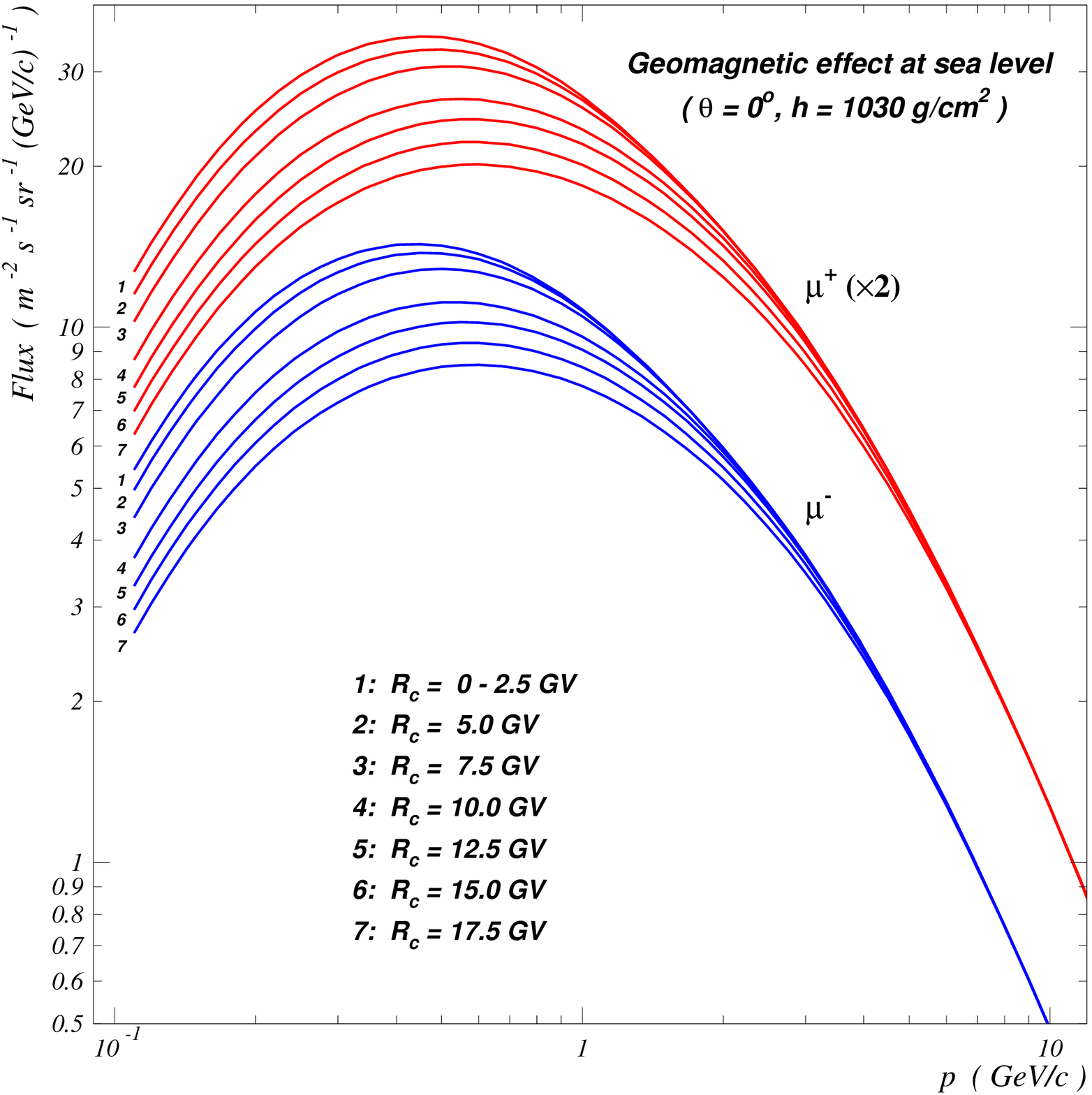} 
\includegraphics[width=8.4cm]{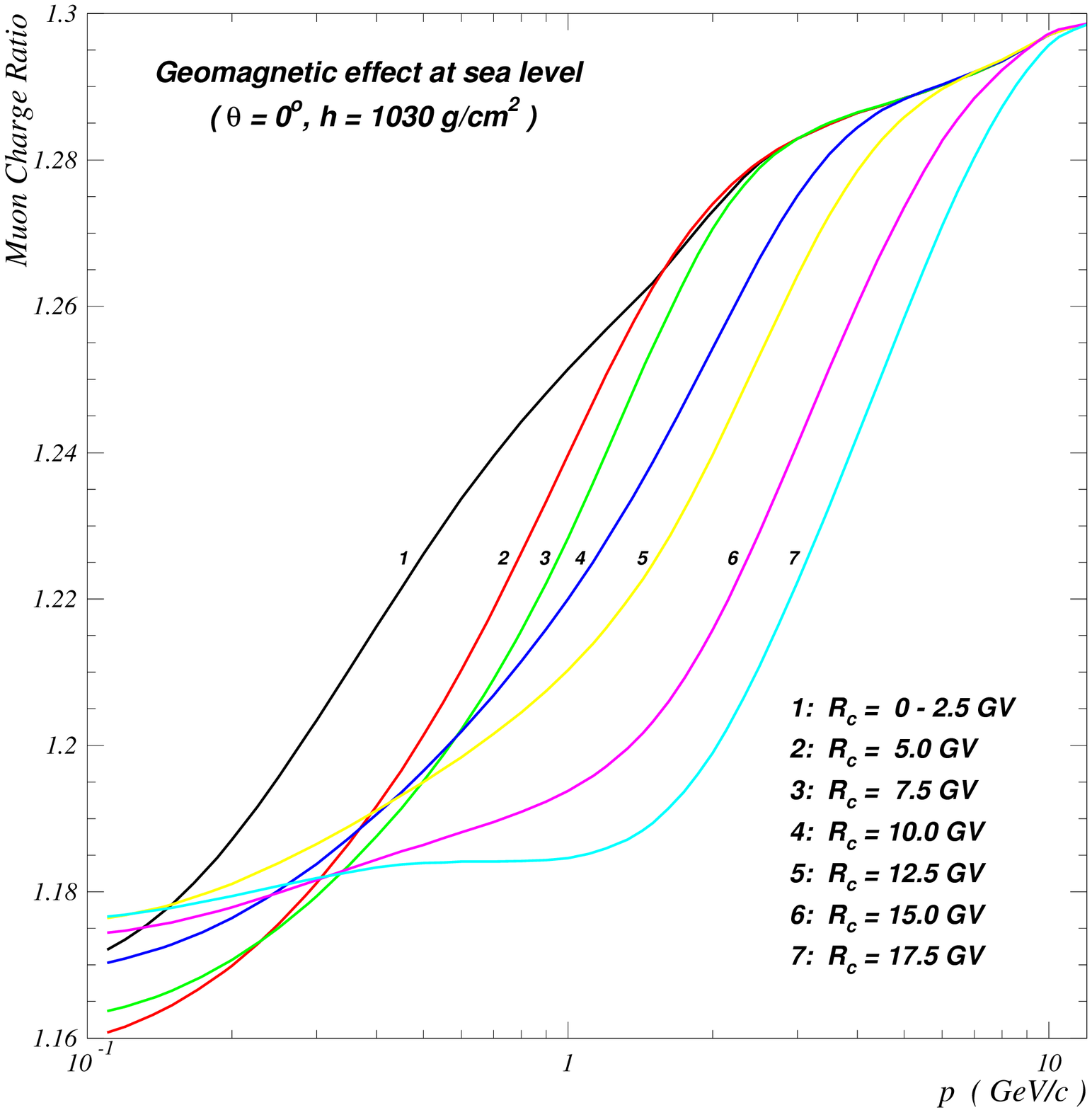} 
\fi
\protect\caption{Vertical differential momentum spectra of positive
                 and negative muons and muon charge ratios at
                 sea level calculated with CORT for several
                 geomagnetic cutoffs $R_c$. The $\mu^+$ spectra on
                 the top panel are scaled with a factor of 2.
\label{f:Geo}}
\end{figure}

Figure \ref{f:Geo} shows expected geomagnetic effect for vertical
differential momentum spectra of $\mu^+$ and $\mu^-$ and for muon
charge ratio $\mu^+/\mu^-$ at sea level. A comparison of these
predictions with experiment will be discussed elsewhere.

\begin{figure*}
\centering
\ifpdf
\includegraphics[width=0.90\linewidth]{fsltot.pdf} 
\includegraphics[width=0.90\linewidth]{fsl-he.pdf} 
\else
\includegraphics[width=0.90\linewidth]{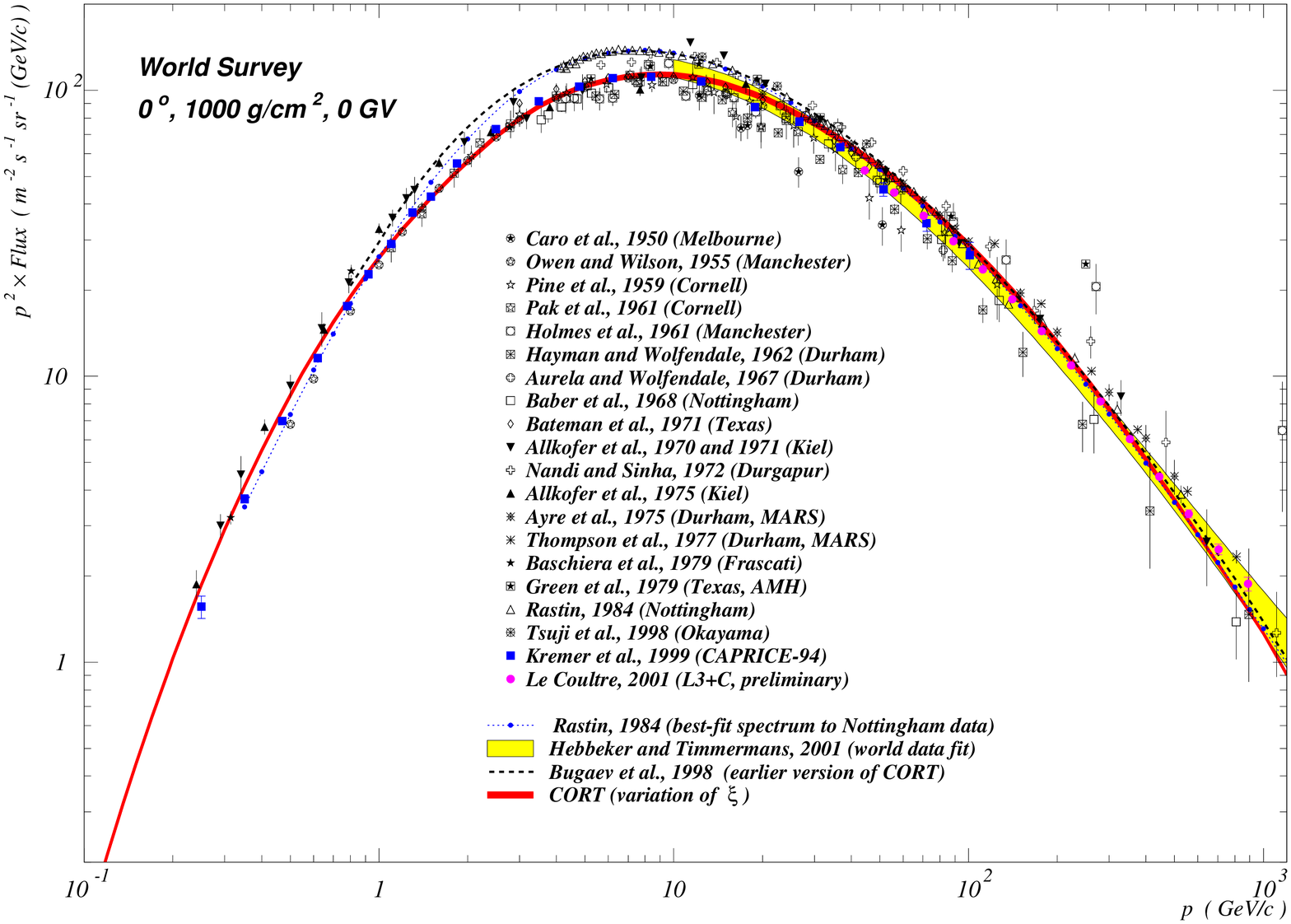} 
\includegraphics[width=0.90\linewidth]{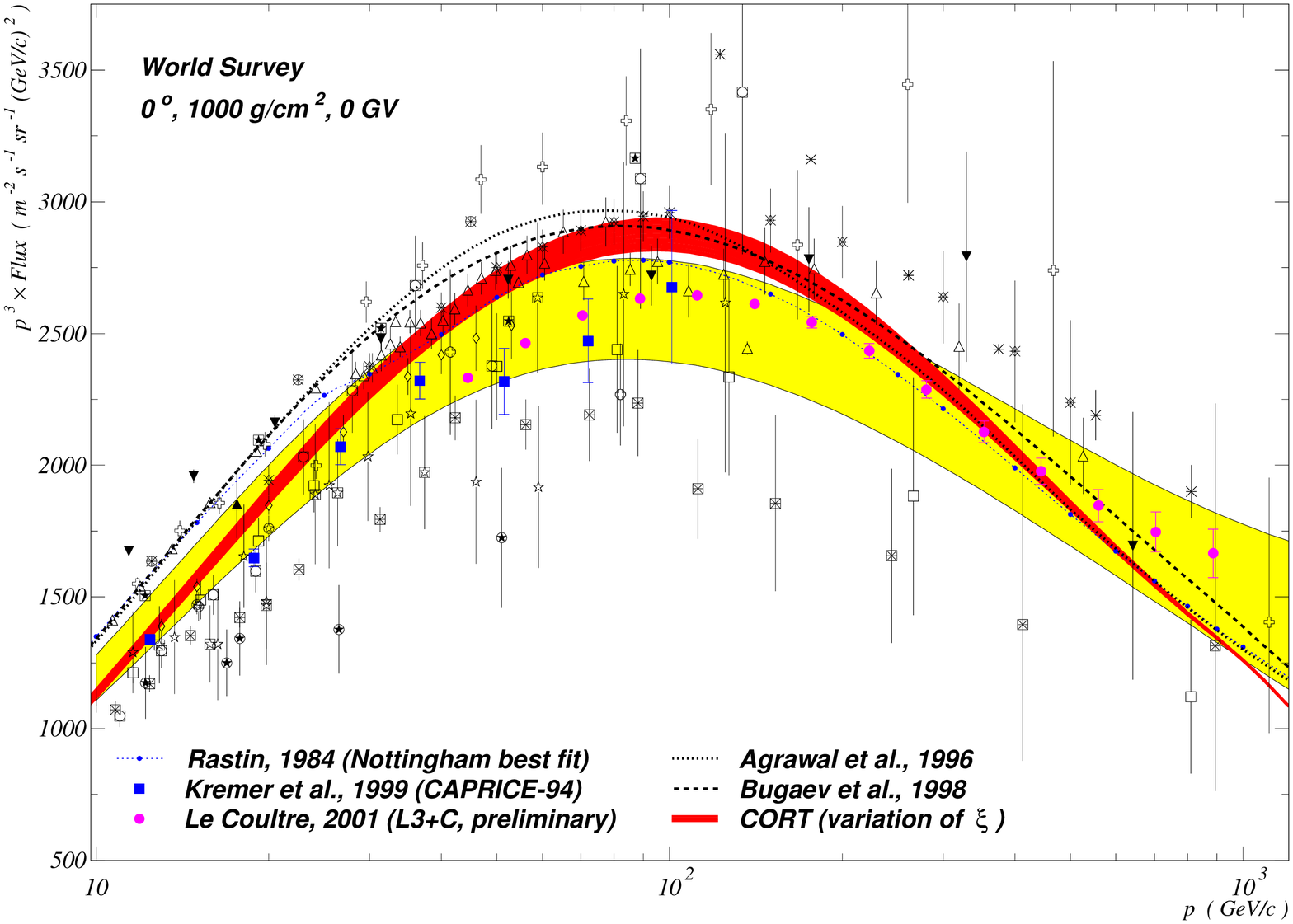} 
\fi
\protect\caption{Near-vertical differential momentum spectrum of
                 $\mu^++\mu^-$ at ground level (top panel) and its
                 rescaled fragment for $p>10$ GeV/c (bottom panel).
                 The thickness of the band calculated with CORT
                 reflects the  indetermination in the parameter
                 $\xi$. 
\label{f:FslSurvey}}
\end{figure*}

Figure \ref{f:FslSurvey} collects the data on the near-vertical
differential momentum spectrum of $\mu^++\mu^-$ at ground level
from many experiments performed over the past five decades
\citep{Caro50,Owen55,Pine59,Pak61,Holmes61,Hayman62a,Hayman62b,%
       Aurela67,Baber68a,Allkofer70a,Allkofer70b,Allkofer71,%
       Bateman71,Nandi72a,Allkofer75,Ayre75,Thompson77,Baschiera79,%
       Green79,Rastin84a,Tsuji98,Kremer99,LeCoultre01}.
Only the data for muon momenta below $\sim1$ TeV/c are shown.

The compilation, of course, is not exhaustive. In particular, it
does not involve the low-energy data for large geomagnetic cutoffs,
the ground-level data obtained at the altitudes far different from
sea level, and also indirect muon data from underground
measurements.%
\footnote{These were discussed in detail by \citet{Bugaev98}.}
The spectra calculated for $h=10^3$ g/cm$^2$, $\theta=0^\circ$,
and $R_c=0$ with the new and old (from \citet{Bugaev98}) versions
of CORT are plotted. The best-fit spectrum to Nottingham data
\citep{Rastin84a} and the spectrum calculated by \citet{Agrawal96}
are also shown in bottom panel.

In fact, not all experiments listed above measure at sea level.
Moreover, the experiments were performed in different locations
and periods. Therefore the solar modulation effects, meteorological
and geomagnetic conditions are generally different in these
experiments.
\begin{figure*}
\centering
\ifpdf
\includegraphics[width=\linewidth]{crsl.pdf} 
\else
\includegraphics[width=\linewidth]{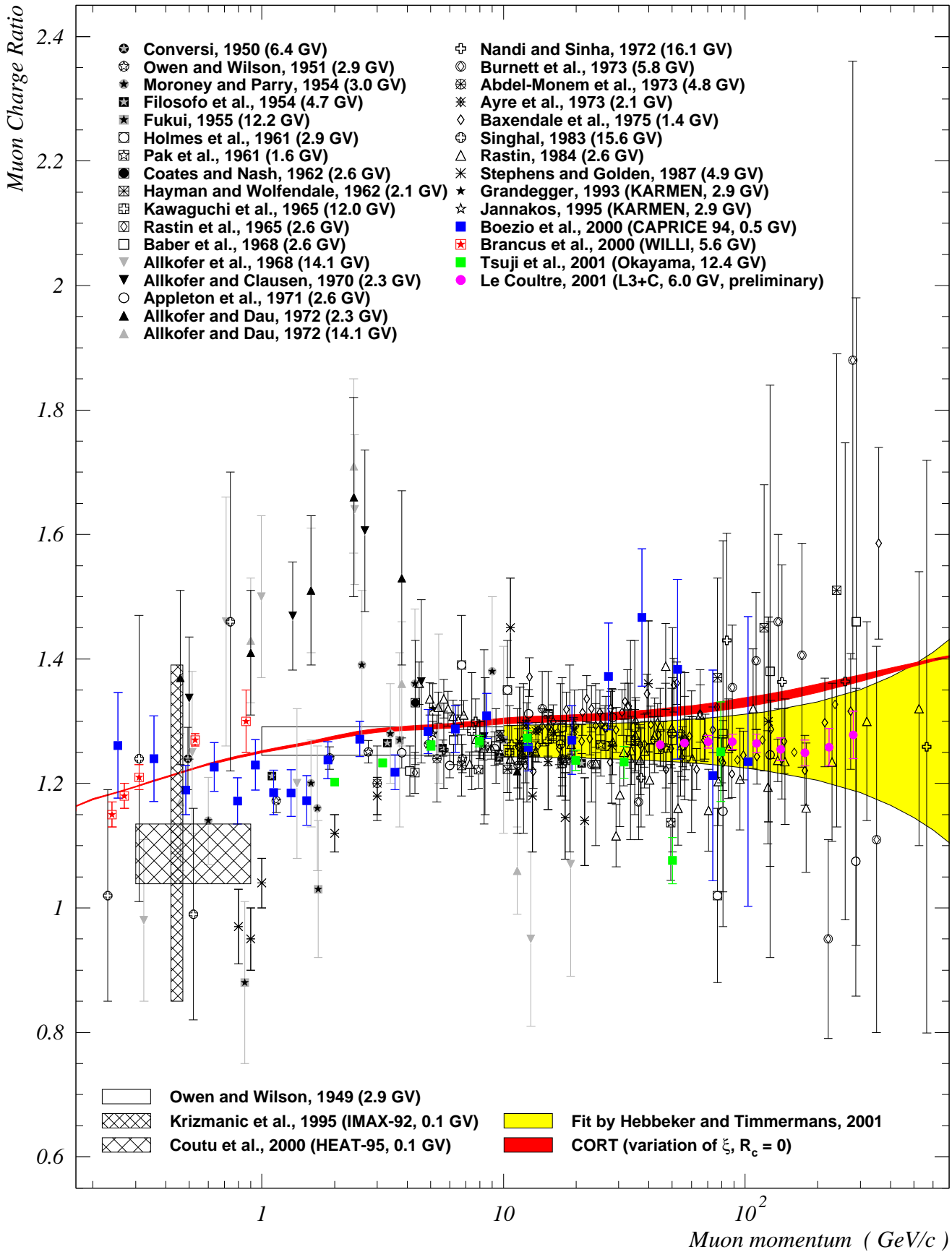} 
\fi
\protect\caption{Muon charge ratio at sea level for near-vertical
                 direction.
                 Geomagnetic cutoffs $R_c$ for different experiments
                 are shown in the legend.
\label{f:CRSL}}
\end{figure*}
The same is true for the detection cones, shielding factors, adopted
procedures for the momentum spectrum unfolding, and so on. All these
factors might cause a bias when all the data are compared together
\citep{Hebbeker01}. Here we do not pursue any correction procedure
(like one suggested by \citet{Hebbeker01}) and plot all the
experimental data as they are. Instead we show the world average fit
derived by \citet{Hebbeker01} from an analysis of (almost) the same
data set at $p\geq10$~GeV/c.
It can be seen that the ground-level experiments are in rather poor
agreement to one another. A large share of the data are arranged
outside (or even far from) the borderline of the world average fit.
However the data of the most recent experiments, specifically
CAPRICE\,94 \citep{Kremer99} and L3+C \citep{LeCoultre01} are
placed close to or within these bounds.%
\footnote{Note that the L3+C data on muon flux and charge ratio
          reported in ICRC'27 \citep{LeCoultre01} (see also Refs.
          \citep{Petersen01,Ladron01}), are preliminary and may be
          subject to refinement (within the systematic error of
          7.7\%) when analyses are complete.}

Fig. \ref{f:FslSurvey} suggests that the muon flux calculated
with the new CORT is systematically lower than that obtained by
\citet{Bugaev98} using the earlier version. The deviation is about
18\% for $p=10$~GeV/c and about a few per cent for
$p\gtrsim80$~GeV/c.
The main reason of this difference is that the BESS+JACEE primary
spectrum below $\sim200$~GeV/nucleon is essentially lower than
that used by \citet{Bugaev98} and in the earlier calculations
with CORT. Another reasons pertain to the changes in the inclusive
cross sections and improved description of muon propagation.
Below $\sim400$ GeV/c the new spectrum is close to the world
average fit and to the data from CAPRICE\,94 and L3+C experiments
as well as to the data from several experiments carried out by
the BESS Collaboration \citep{Sanuki01a,Motoki01,Sanuki01b}.%
\footnote{Unfortunately, the tabulated muon data are still
          unavailable from the BESS Collaboration.}

Let us now dwell to the muon charge ratio. It has been shown
\citep{Fiorentini01a} that calculations with CORT describe well
the modern data on the charge ratio obtained at different atmospheric
depths. Here we consider only the ground-level data for near-vertical
directions. The world survey is presented in Fig. \ref{f:CRSL}.
The data are taken from many experiments
\citep{Owen49,Conversi50,Owen51,Moroney54,Filosofo54,Fukui55,%
       Holmes61,Pak61,Coates62,Hayman62a,Hayman62b,Kawaguchi65,%
       Rastin65,Baber68b,Allkofer68,Allkofer70a,Appleton71,%
       Allkofer72,Nandi72b,Burnett73,Abdel-Monem73,Ayre73,%
       Baxendale75,Singhal83,Rastin84b,Stephens87,Grandegger93,%
       Jannakos95,Krizmanic95,Boezio00,Coutu00,Brancus00,%
       Tsuji01,LeCoultre01},
performed at different geomagnetic locations. According to CORT (see
bottom panel of Fig. \ref{f:Geo}), the sea-level charge ratio is
slightly affected by the geomagnetic field. However the dispersion
of the experimental points is too large and they do not show some
significant correlation with the geomagnetic cutoff.
\begin{figure}[hbt]
\centering
\ifpdf
\includegraphics[width=8.3cm]{amh.pdf} 
\else
\includegraphics[width=8.3cm]{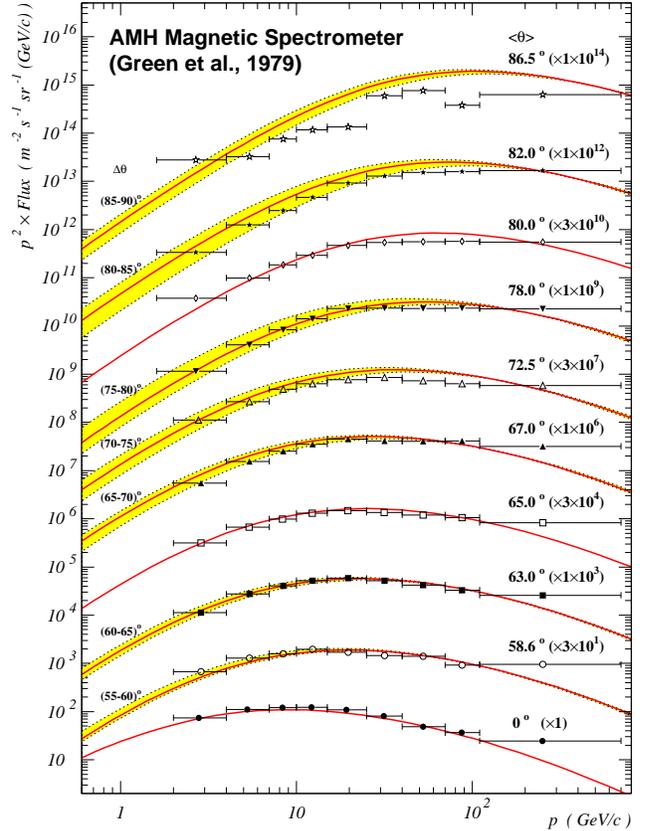} 
\fi
\protect\caption{Differential momentum spectra of muons at sea level
                 for several zenith angles and angular bins. The data
                 points are from AMH spectrometer \citep{Green79}.
                 The filled areas display the expected variations of
                 the muon fluxes inside the bins
                 \protect$\Delta\theta$ indicated at the left.
                 The solid curves are calculated for average zenith
                 angles \protect$\langle\theta\rangle$. All the data
                 are scaled with the factors shown in the parentheses
                 at the right.
\label{f:AMH}}
\end{figure}
\begin{figure*}[bt]
\centering
\ifpdf
\includegraphics[width=\gw]{f30.pdf}~
\includegraphics[width=\gw]{f75.pdf} 
\else
\includegraphics[width=\gw]{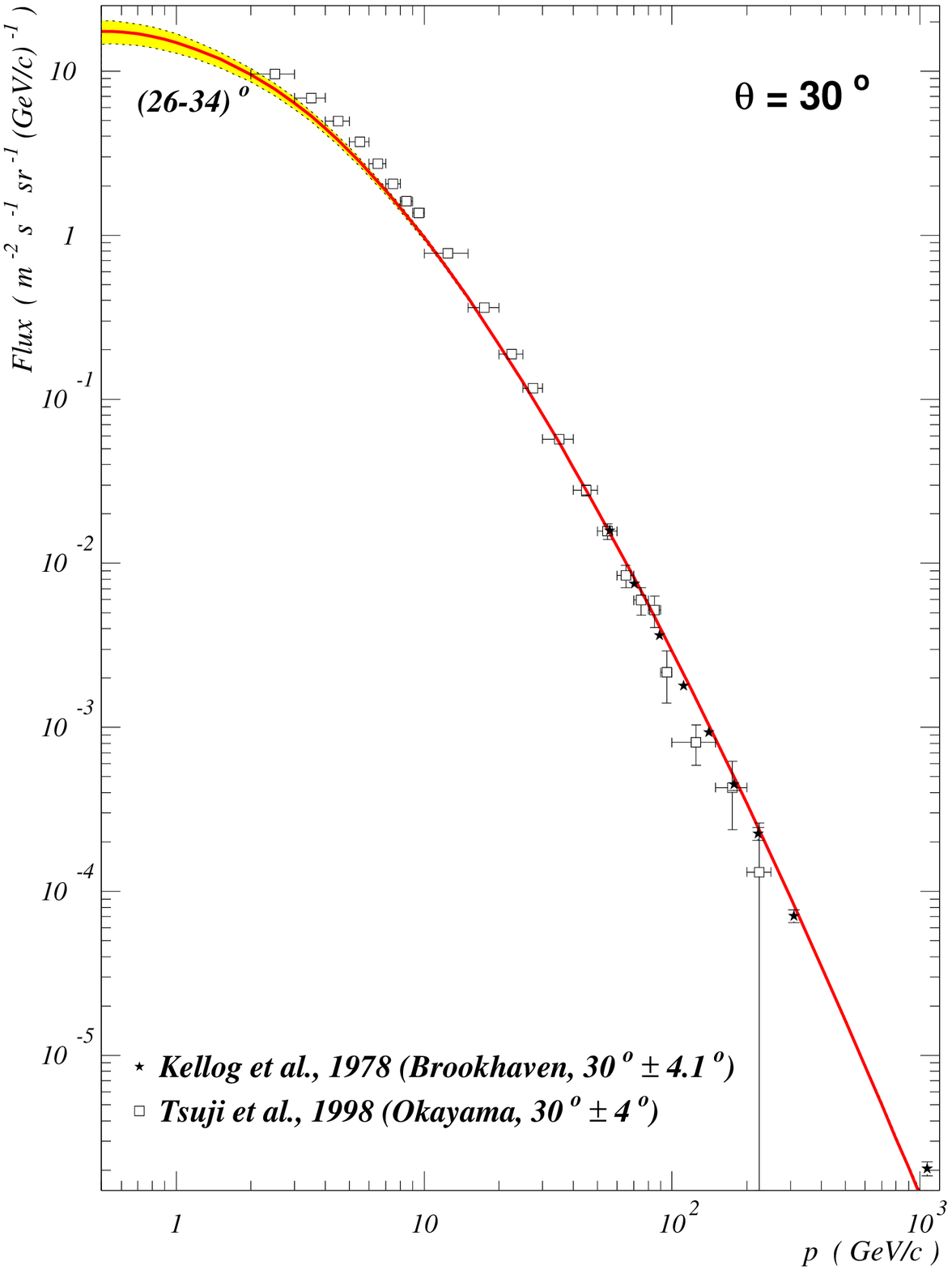}~
\includegraphics[width=\gw]{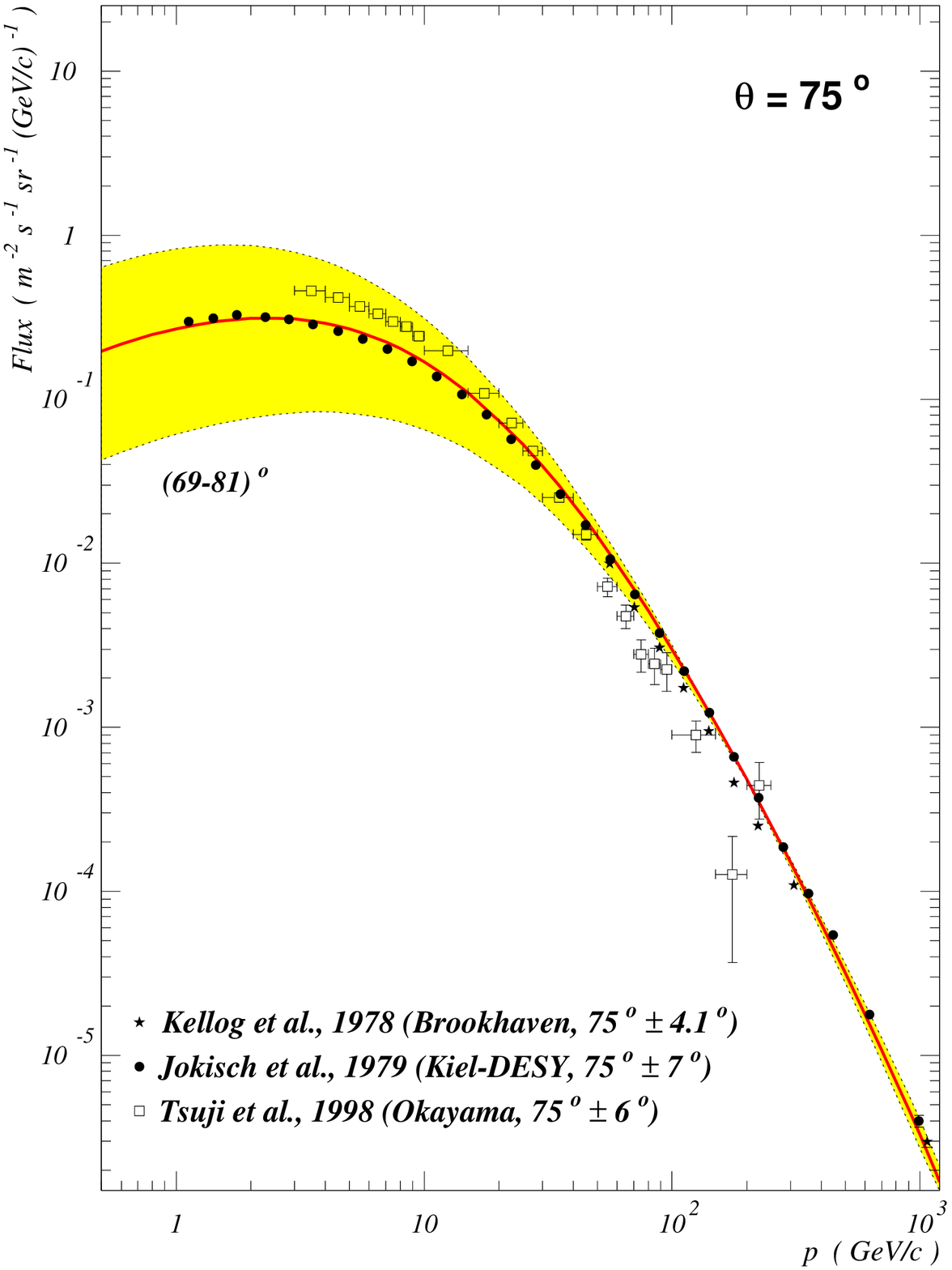} 
\fi
\protect\caption{Differential momentum spectra of muons at sea level
                 for $\theta=30^\circ$ and $75^\circ$. The data
                 points are from \citep{Kellogg78,Jokisch79,Tsuji98}.
                 The filled areas display expected variations of the
                 muon fluxes inside the indicated angular bins.
                 The solid curves are for the corresponding
                 average zenith angles.
\label{f:F30-75}}
\end{figure*}

Figure \ref{f:AMH} shows a comparison with the sea-level data from
the AMH magnetic spectrometer (Texas A \& M and University of
Houston Collaboration) measured within different zenith-angle bins
$\Delta\theta$ \citep{Green79}. The spectra for average zenith
angles and expected variations of the spectra inside the
angular bins are shown by solid lines and filed areas, respectively.
One can see a good or at least qualitative agreement with the data
for $p\lesssim100$ GeV/c and $\theta\lesssim80^\circ$. At higher
momenta and at large zenith angles the situation is spoiled.
However, a comparison between the AMH and world survey data
for vertical (see Fig.~\ref{f:FslSurvey}) suggests that there
is some systematic bias in the AMH experiment above 100 GeV/c.
The abnormal scatter of points in the near-horizontal bin
is indicative of a flaw in the large-angle measurements.

Figure \ref{f:F30-75} displays a comparison between the differential
momentum spectra of muons at ground level calculated with CORT and
the data from Brookhaven magnetic spectrometer \citep{Kellogg78},
Kiel-DESY muon spectrometer located at Hamburg \citep{Jokisch79},
and Okayama altazimuthal counter cosmic-ray telescope with a magnet
spectrometer \citep{Tsuji98}.
The filled areas display the expected variations of the muon
fluxes inside the angular bins $\Delta\theta=(26-34)^\circ$
(left panel) and $(69-81)^\circ$ (right panel).
The solid curves are for the corresponding average zenith angles
of $30^\circ$ and $75^\circ$. The calculations presented in Fig.
\ref{f:F30-75} are done without taking into account the geomagnetic
effects (both the primary spectrum cutoff and muon deflection).
These are not important under the conditions of Brookhaven and
Kiel-DESY experiments but requisite for Okayama site.%
\footnote{The Okayama telescope location is
          ($34^\circ40'$\,N,\,$133^\circ56'$\,E) and vertical
          geomagnetic cutoff rigidity is about 12.4 GV.}

Generally the geomagnetic effects must decrease the low-momentum part
of the expected muon flux (see top panel of Fig. \ref{f:Geo}). The
calculated spectra fit rather well the precise Brookhaven and
Kiel-DESY data but are in rather bad agreement (especially for
$\theta=30^\circ$) with the Okayama telescope data of poorer
statistics; the geomagnetic corrections can only aggravate
the disagreement. One can conclude from Fig. \ref{f:F30-75} that
there is a systematic flaw in the average zenith angles measured
in the Okayama experiment.

\protect\section{Numerical results for neutrinos}
\label{sec:Neutrinos}

\begin{figure}[thb]
\centering
\ifpdf
\includegraphics[width=\linewidth]{a.pdf} 
\else
\includegraphics[width=\linewidth]{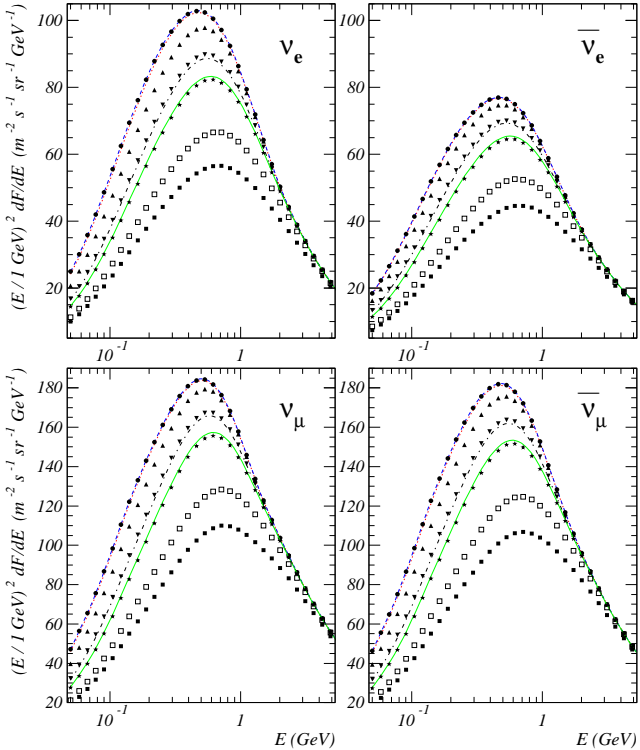} 
\fi
\protect\caption{Scaled $4\pi$ averaged fluxes of $\nu_e$,
                 $\overline{\nu}_e$, $\nu_\mu$, and
                 $\overline{\nu}_\mu$ for ten underground
                 laboratories (see table \ref{t:Labs} for notation).
\label{f:ANFluxLE}}
\end{figure}

\begin{figure}[thb]
\centering
\ifpdf
\includegraphics[width=\linewidth]{r.pdf} 
\else
\includegraphics[width=\linewidth]{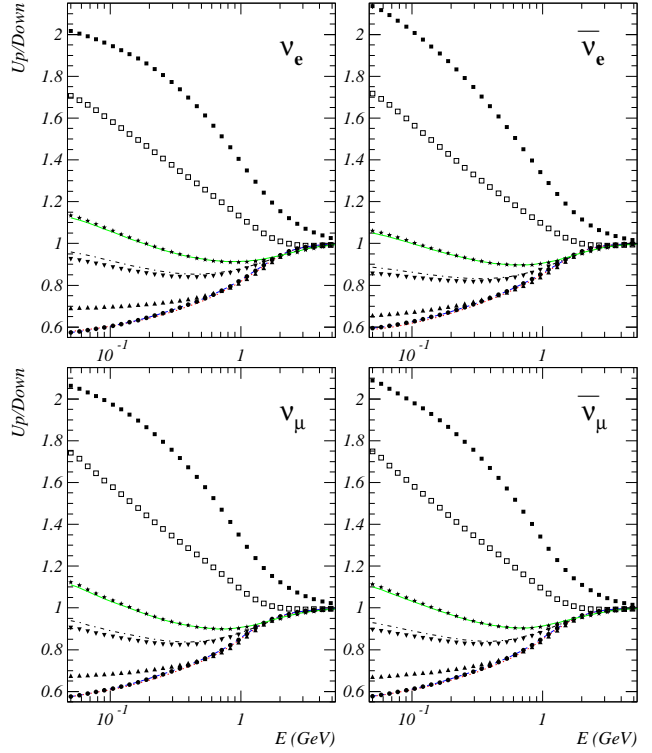} 
\fi
\protect\caption{Up-to-Down ratios of the $\nu_e$,
                 $\overline{\nu}_e$, $\nu_\mu$, and
                 $\overline{\nu}_\mu$ fluxes for ten underground
                 laboratories (see table \ref{t:Labs} for notation).
\label{f:ANRatioLE}}
\end{figure}

Let us briefly sketch some results on atmospheric neutrinos (AN).
Due to geomagnetic effects, the low-energy AN spectra and angular
distributions are quite different for different sites of the globe.
Figures \ref{f:ANFluxLE} and \ref{f:ANRatioLE} display our
predictions for ten underground neutrino laboratories listed in
Table \ref{t:Labs}.
\begin{table}[htb]
\centering
\protect\caption{List of some underground laboratories.
\label{t:Labs}}
\vspace{3mm}
\ifpdf
\includegraphics[width=\linewidth]{labs.pdf} 
\else
\includegraphics[width=\linewidth]{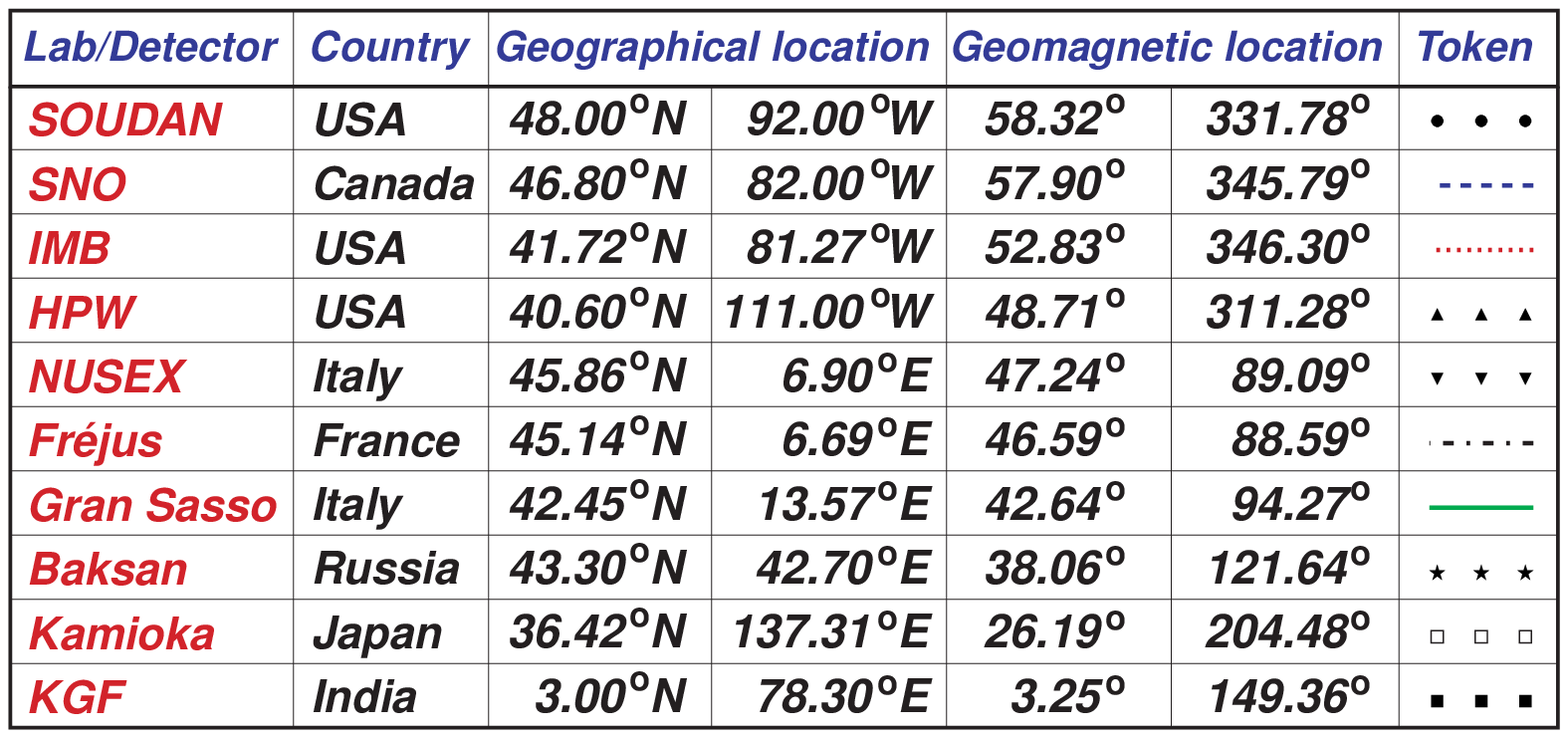} 
\fi
\end{table}

Figure \ref{f:ANFluxLE} shows the $\nu_e$, $\overline{\nu}_e$,
$\nu_\mu$ and $\overline{\nu}_\mu$ energy spectra averaged over all
zenith and azimuth angles. The ratios of the AN fluxes averaged over
the lower and upper semispheres (``up-to-down'' ratios) are shown in
Fig. \ref{f:ANRatioLE}. As a result of geomagnetic effects, both the
spectra and the up-to-down ratios for the following 6 groups of the
labs: SOUDAN + SNO + IMB, HPW, NUSEX + Fr\'ejus, Gran Sasso + Baksan,
Kamioka, and KGF are quite distinct for energies below a few GeV.

\begin{figure*}[t]
\centering
\ifpdf
\includegraphics[width=\w]{e1k.pdf}~
\includegraphics[width=\w]{e2k.pdf}~
\includegraphics[width=\w]{e1g.pdf}~
\includegraphics[width=\w]{e2g.pdf} 
\includegraphics[width=\w]{m1k.pdf}~
\includegraphics[width=\w]{m2k.pdf}~
\includegraphics[width=\w]{m1g.pdf}~
\includegraphics[width=\w]{m2g.pdf} 
\else
\includegraphics[width=\w]{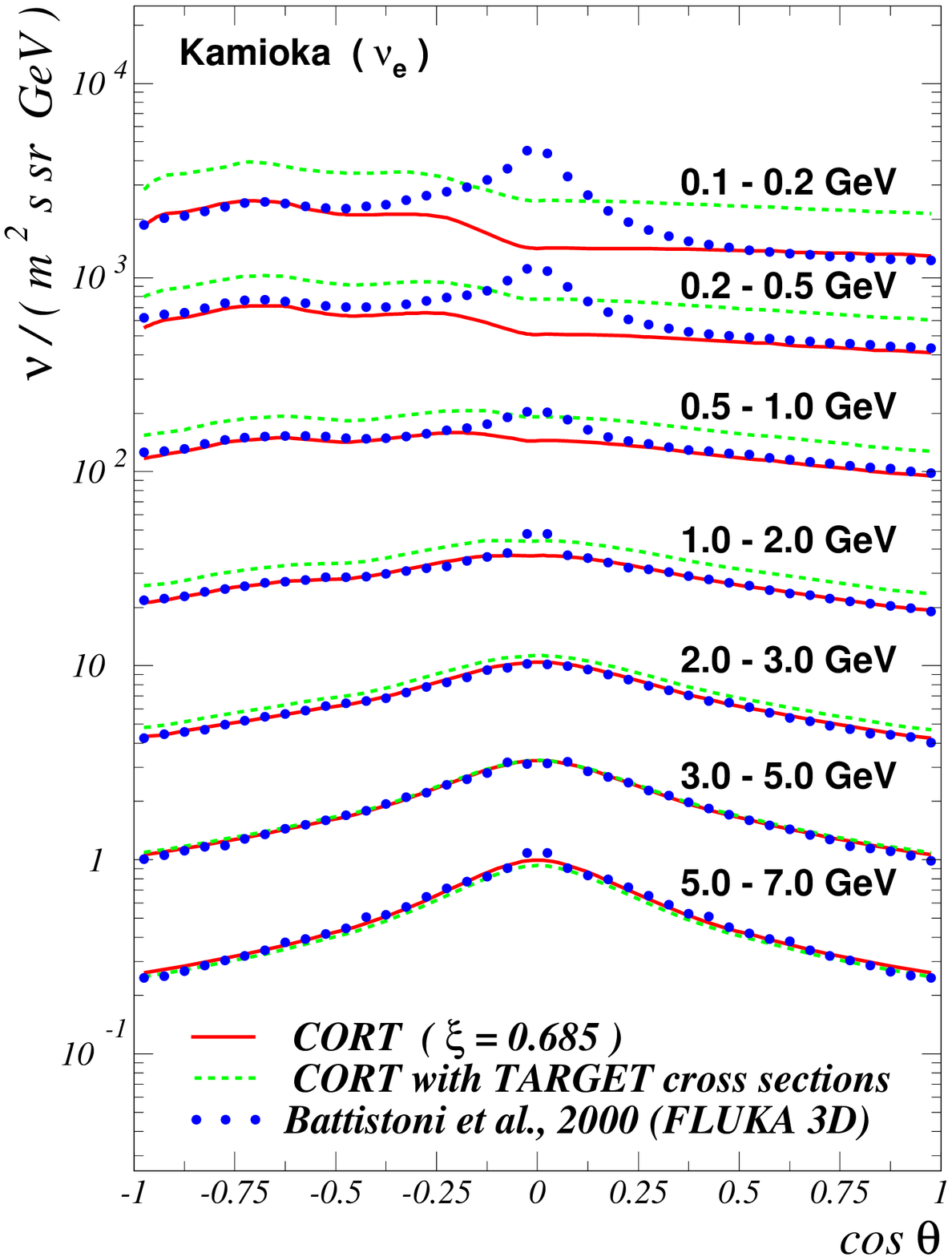}~
\includegraphics[width=\w]{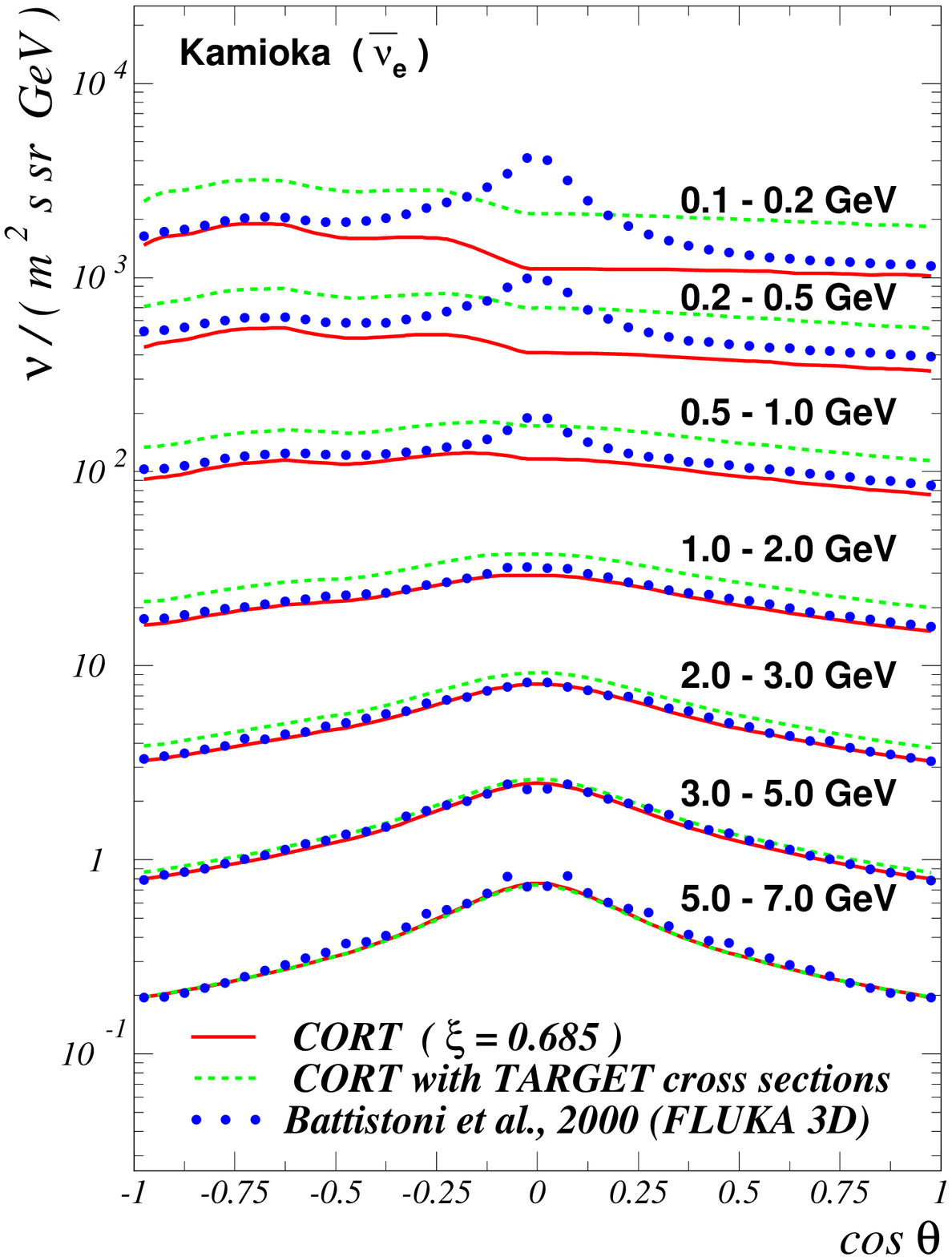}~
\includegraphics[width=\w]{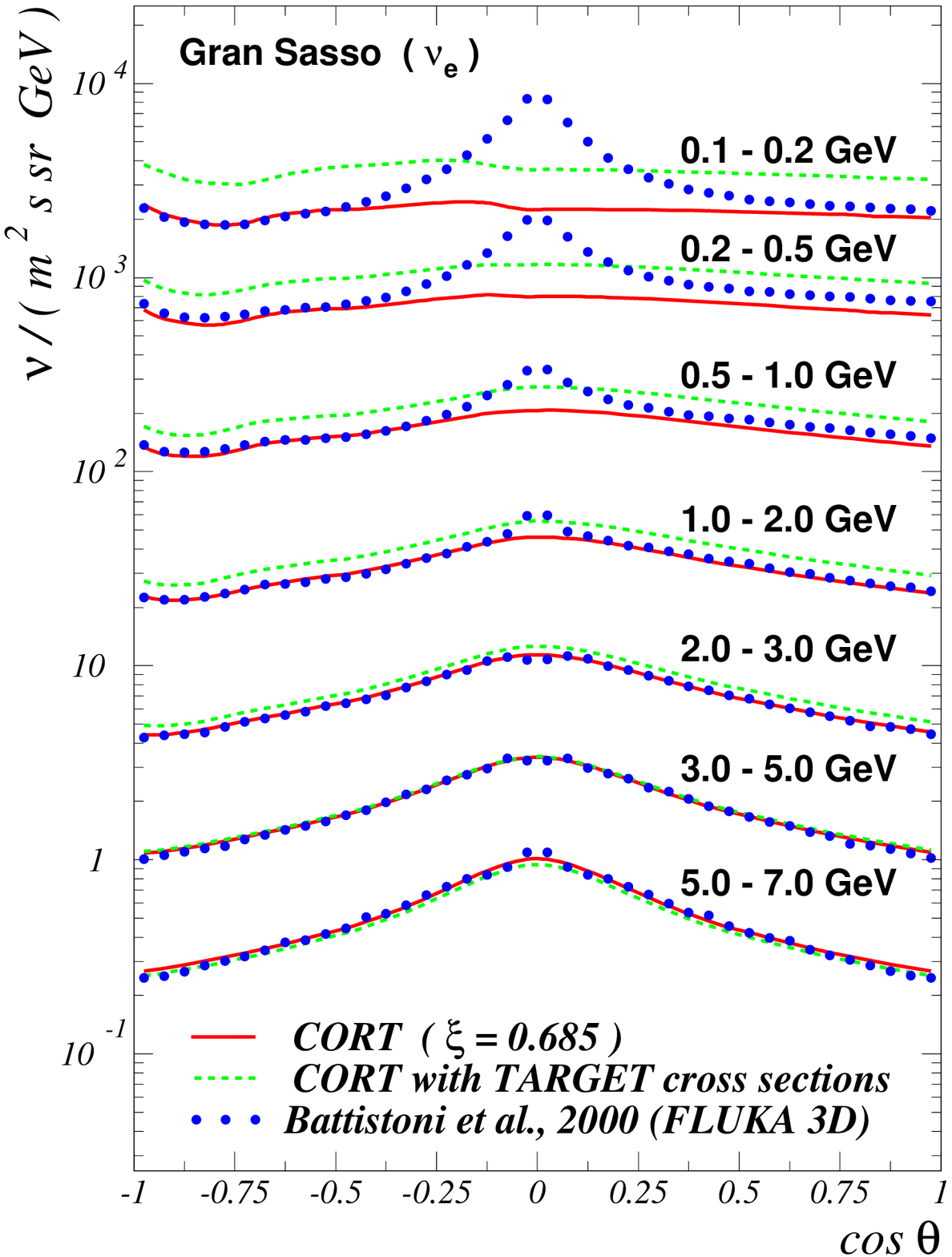}~
\includegraphics[width=\w]{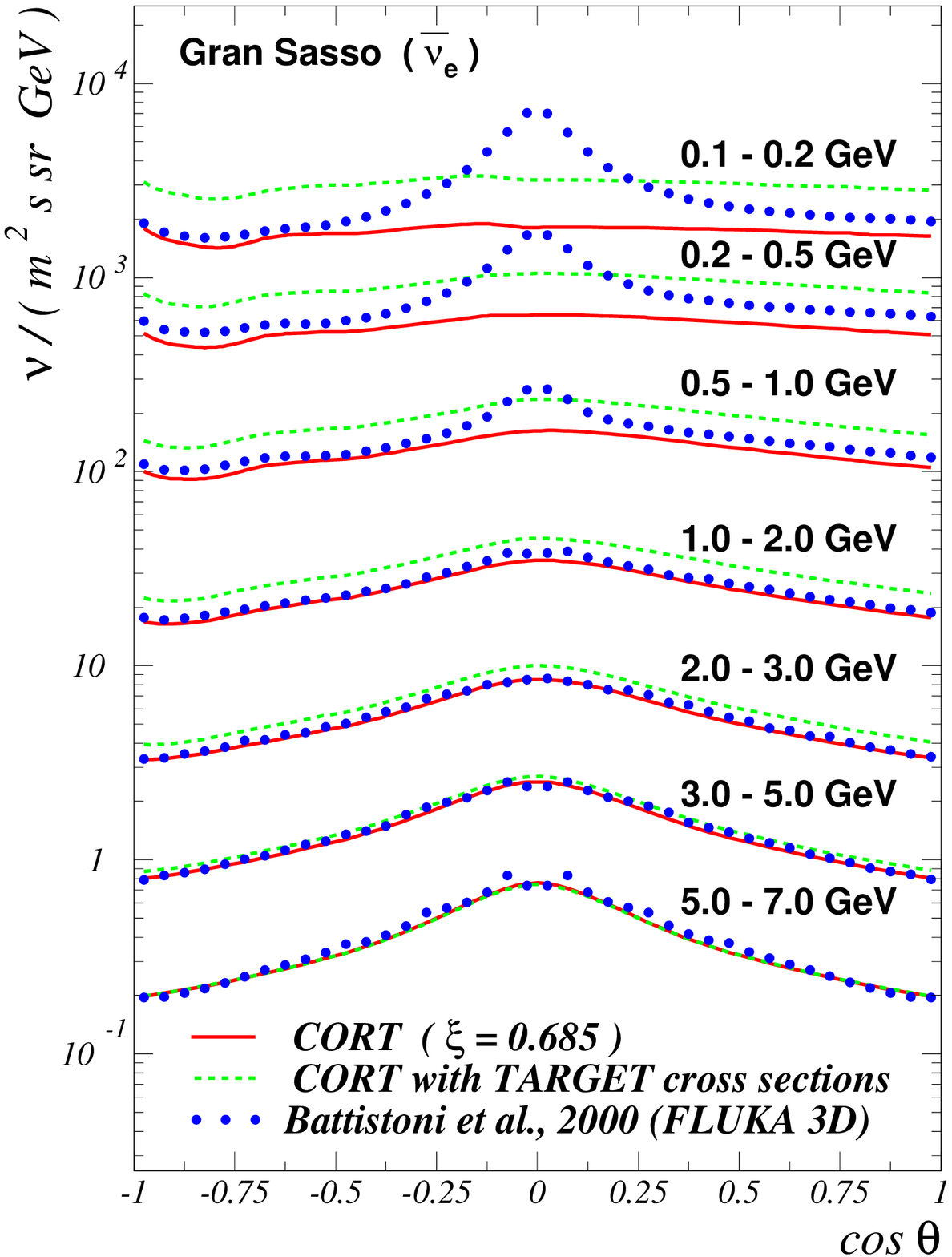} 
\includegraphics[width=\w]{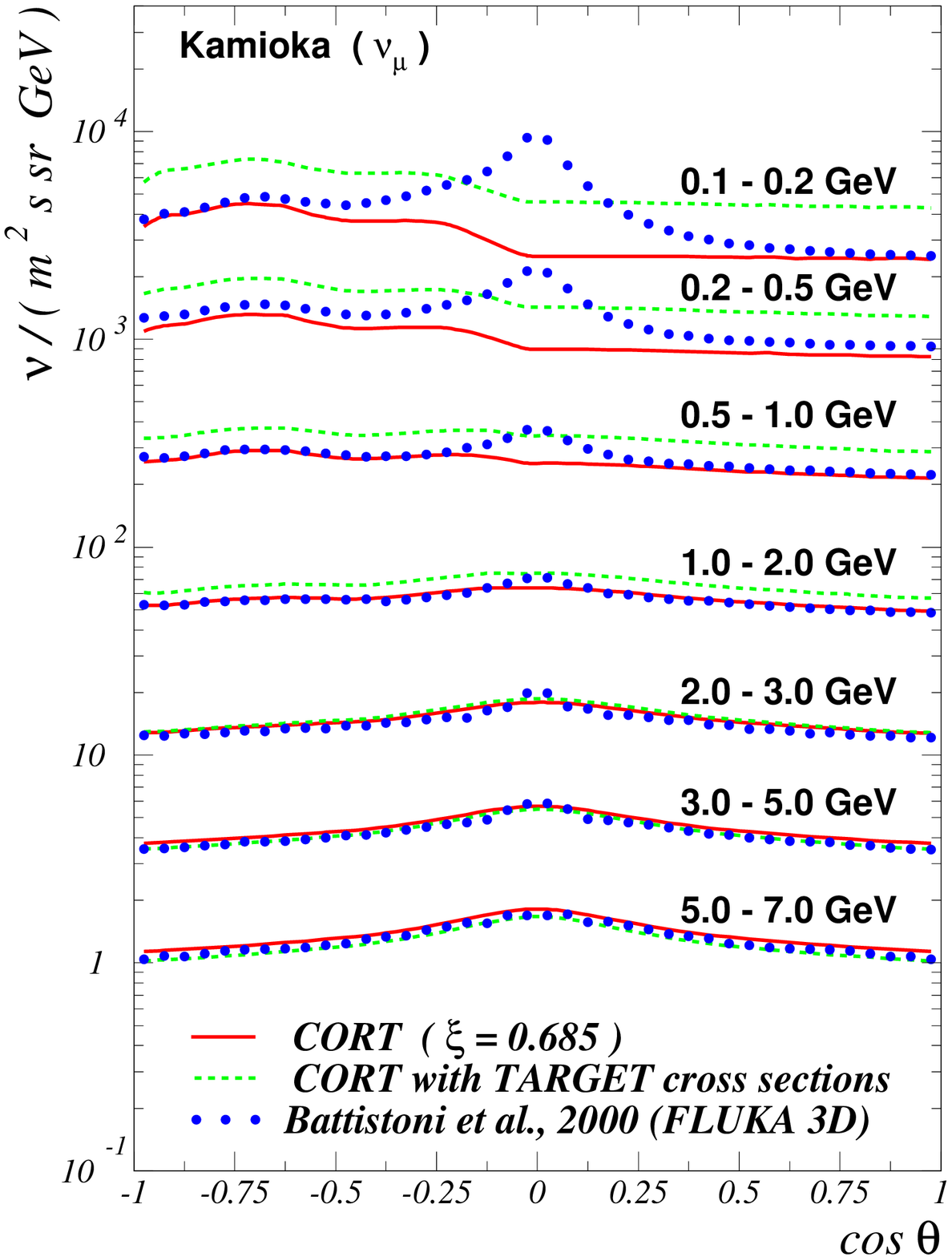}~
\includegraphics[width=\w]{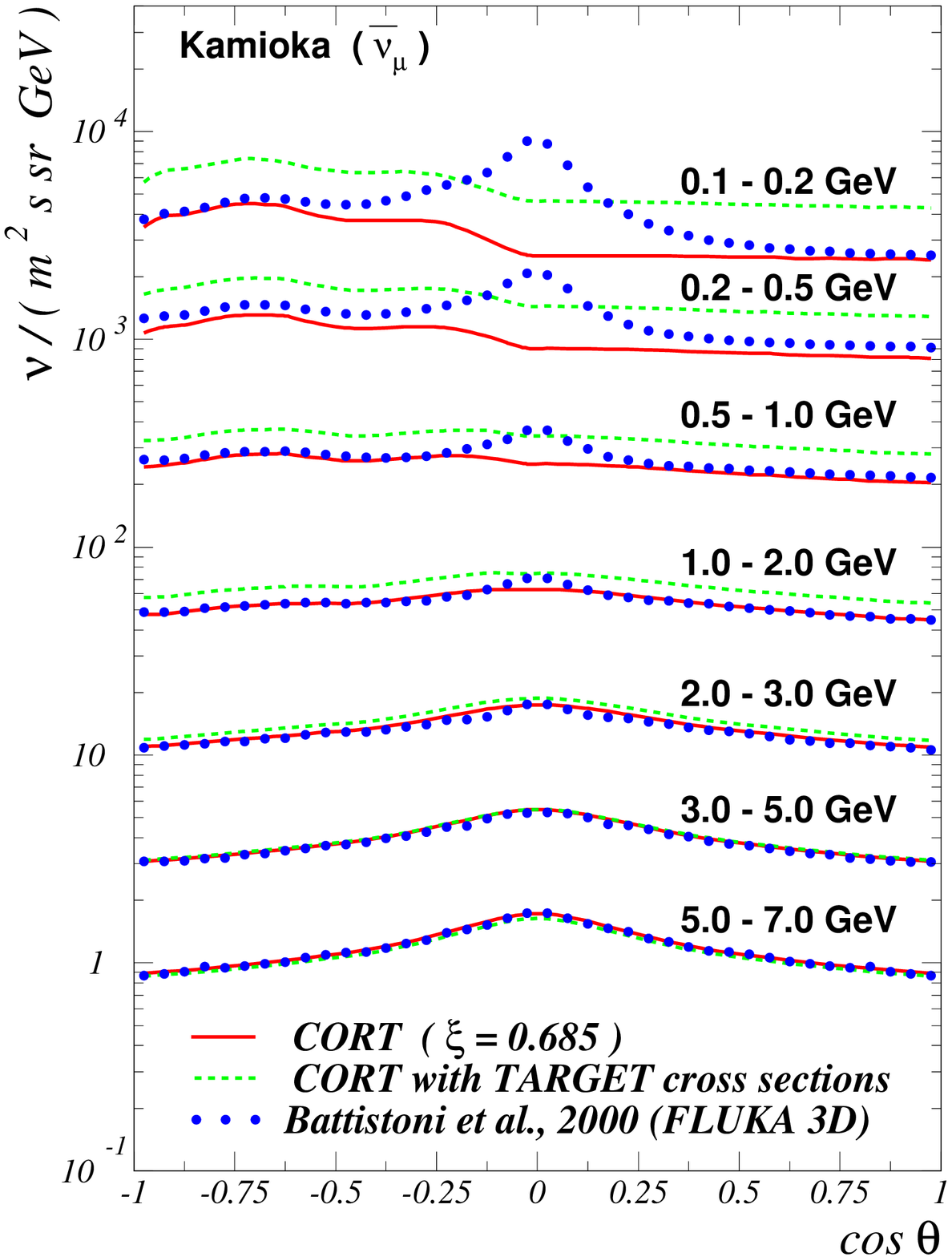}~
\includegraphics[width=\w]{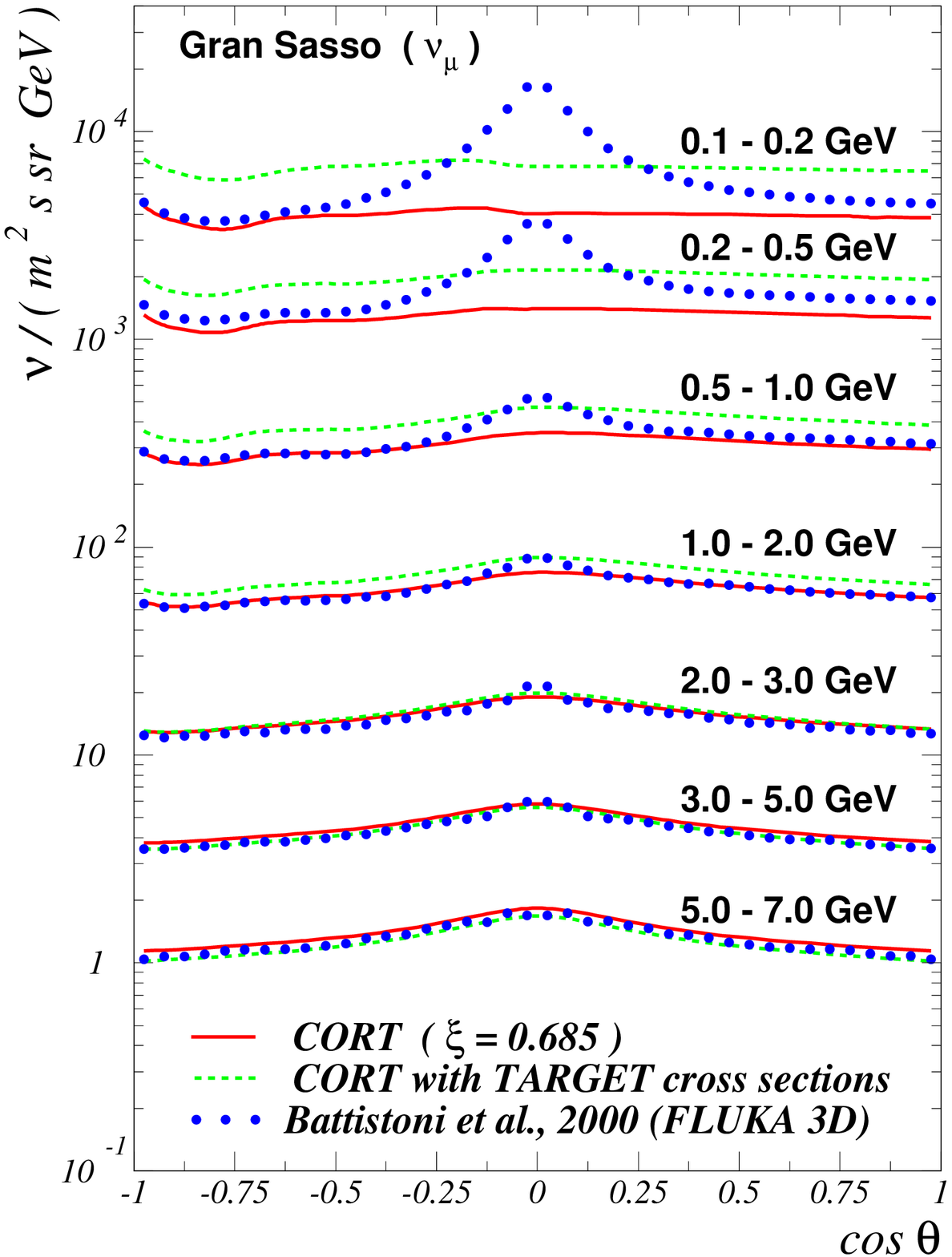}~
\includegraphics[width=\w]{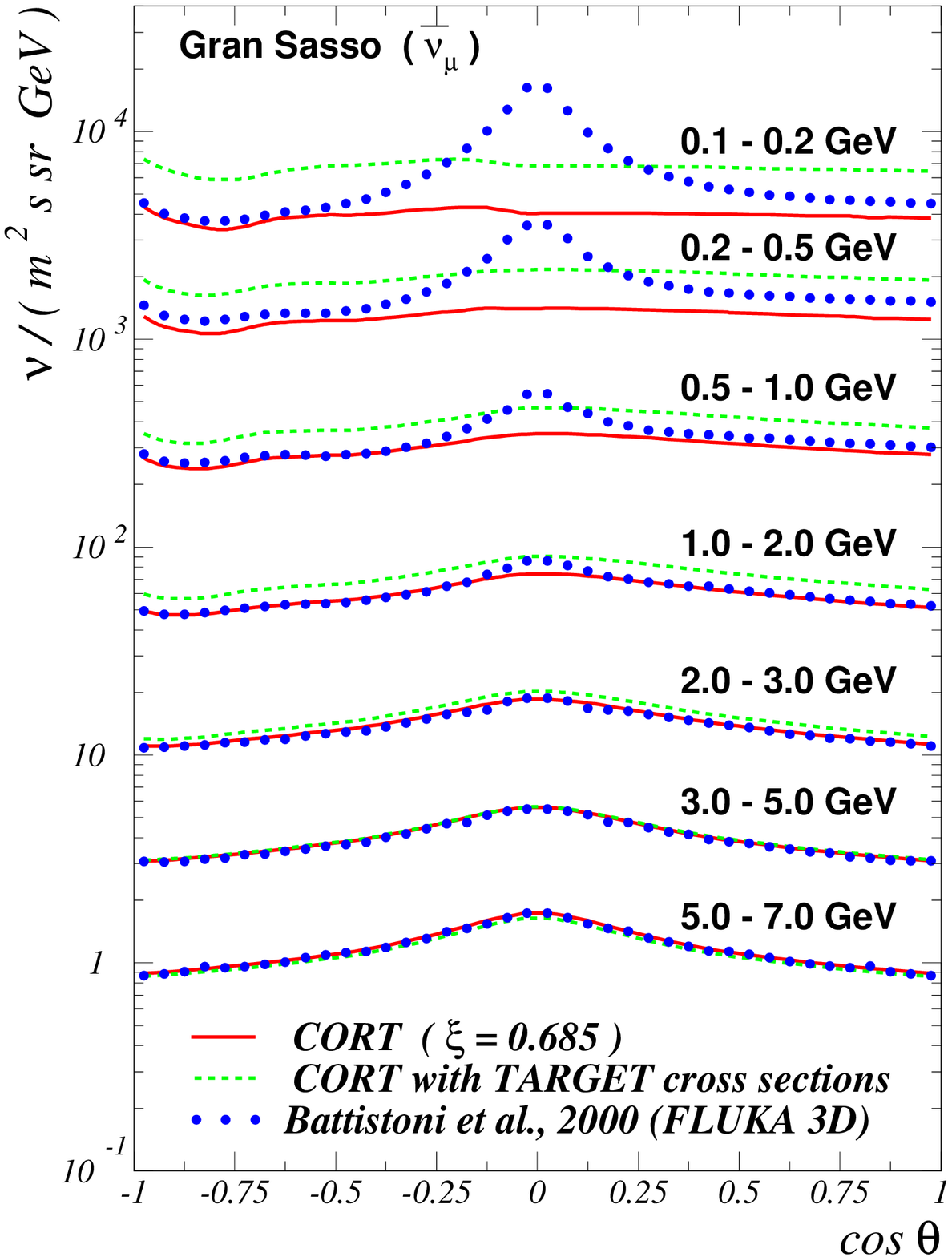} 
\fi
\protect\caption{Zenith-angle distributions of $\nu_e$,
                 $\overline{\nu}_e$, $\nu_\mu$, and
                 $\overline{\nu}_\mu$ for Kamioka and Gran Sasso.
                 Solid curves represent the result of CORT obtained
                 with its ``standard'' interaction model.
                 The dashed curves are the result of CORT obtained
                 with the TARGET model for meson production and
                 superposition model for collisions of nuclei.
                 The circles are for the result of calculation
                 by \citet{Battistoni00} based on FLUKA 3D code.
                 The distributions are averaged over the azimuth
                 angle and over the energy bins indicated near the
                 curves.
\label{f:K-GS}}
\end{figure*}

Figure \ref{f:K-GS} depicts the zenith-angle distributions of
$\nu_e$, $\overline{\nu}_e$, $\nu_\mu$ and $\overline{\nu}_\mu$
for Kamioka and Gran Sasso labs calculated with CORT using its
``standard''(KM+SS) model of particle interaction
(see sec. \ref{sec:interactions}) and also the TARGET model for
$\pi/K$ meson production (including the superposition model for
collisions of nuclei) used by Bartol group
\citep{Barr89,Agrawal96,Lipari98}.
The TARGET model for nucleon production is not included into the
CORT options yet. Instead the KM+SS model is used everywhere.
Below, we refer to CORT switched into the
TARGET meson production model as ``CORT+TARGET''.
For comparison, the result of the recent 3D calculation by
\citet{Battistoni00} based on the FLUKA 3D Monte Carlo simulation
package is also shown. It allows to ``highlight'' the 3D effects
which are very dependent on neutrino energy and direction of
arrival.

A more quantitative comparison is given by Table \ref{t:FluxRatios}
which tabulates the ratios of the$\nu_e$, $\overline{\nu}_e$,
$\nu_\mu$, and $\overline{\nu}_\mu$ fluxes calculated with
CORT+TARGET and FLUKA to those with the ``standard'' CORT.
The fluxes are averaged over all directions and over several
energy bins. Table \ref{t:FluxRatios} also shows the
so-called flavor ratio%
\footnote{This quantity roughly represents the ratio of $e$
          like to $\mu$ like single-ring contained events measured
          in water Cherenkov detectors and the ``showers-to-tracks''
          ratio measured in iron detectors.}
\[
R_\nu=\left(\nu_e  +\tfrac{1}{3}\overline{\nu}_e  \right)/
      \left(\nu_\mu+\tfrac{1}{3}\overline{\nu}_\mu\right),
\]
evaluated with CORT, CORT+TARGET, and FLUKA for the same energy
bins. 

By using Fig. \ref{f:K-GS} and Table \ref{t:FluxRatios}, one can
conclude that the present calculations with CORT for neutrino
energies below 1-2 GeV lead to the AN fluxes which are essentially
{\em lower} than those obtained by \citet{Barr89,Agrawal96,Lipari98}
and by \citet{Honda90,Honda95,Honda96} and those are in current use
for many analyses of sub-GeV and multi-GeV $\nu$ induced events in
underground detectors.%
\footnote{A comparison between the earlier AN flux calculations
          has been discussed in detail by \citet{Gaisser96}.}
On the other hand, the new AN fluxes are rather close to the
those obtained with the earlier versions of CORT
\citep{Bugaev87a,Bugaev87b,Bugaev89a,Bugaev90}. The main
differences (for neutrino energies above 1-2 GeV) are due to the
new data for the primary cosmic-ray spectrum and
composition (see Section \ref{sec:Primaries}).

\clearpage 
\begin{table*}[b]
\centering
\protect\caption{The ratios of $4\pi$ averaged AN fluxes obtained
                 with CORT+TARGET and FLUKA to those with CORT and
                 neutrino flavor ratios $R_\nu$ calculated with CORT,
                 CORT+TARGET, and FLUKA for Kamioka.}
\label{t:FluxRatios}
\vspace{4mm}
\begin{tabular}{|c|cccc|cccc|}
\hline
               & \multicolumn{8}{c|}%
{{\bf\Red Average fluxes normalized to CORT\Black}}                 \\
\hline
$\Delta E_\nu$ & \multicolumn{4}{c|}{CORT+TARGET}
               & \multicolumn{4}{c|}{FLUKA 3D}                      \\
  (GeV)        &
 $\nu_e$    & $\overline{\nu}_e$ & $\nu_\mu$ & $\overline{\nu}_\mu$ &
 $\nu_e$    & $\overline{\nu}_e$ & $\nu_\mu$ & $\overline{\nu}_\mu$ \\
\hline      &      &      &      &      &      &      &      &      \\
  0.1--0.2  & 1.64 & 1.78 & 1.72 & 1.73 & 1.25 & 1.45 & 1.41 & 1.40 \\
  0.2--0.3  & 1.51 & 1.68 & 1.60 & 1.61 & 1.23 & 1.38 & 1.34 & 1.33 \\
  0.3--0.5  & 1.42 & 1.61 & 1.48 & 1.50 & 1.17 & 1.28 & 1.22 & 1.22 \\
  0.5--0.7  & 1.34 & 1.50 & 1.36 & 1.38 & 1.10 & 1.19 & 1.11 & 1.12 \\
  0.7--1.0  & 1.28 & 1.41 & 1.27 & 1.30 & 1.05 & 1.12 & 1.04 & 1.05 \\
  1.0--2.0  & 1.20 & 1.30 & 1.17 & 1.20 & 1.02 & 1.05 & 0.99 & 1.01 \\
  2.0--3.0  & 1.10 & 1.17 & 1.03 & 1.08 & 0.98 & 1.03 & 0.96 & 0.97 \\
  3.0--5.0  & 1.02 & 1.07 & 0.95 & 1.01 & 0.99 & 1.02 & 0.95 & 0.97 \\
  5.0--7.0  & 0.95 & 0.99 & 0.91 & 0.96 & 1.01 & 1.05 & 0.94 & 0.99 \\
  7.0--10.0 & 0.91 & 0.95 & 0.89 & 0.93 & 1.03 & 1.09 & 0.95 & 1.04 \\
 10.0--20.0 & 0.87 & 0.91 & 0.88 & 0.91 & 1.07 & 1.12 & 0.96 & 1.02 \\
 20.0--30.0 & 0.83 & 0.87 & 0.87 & 0.89 & 1.10 & 1.15 & 0.94 & 1.02 \\
            &      &      &      &      &      &      &      &      \\
\hline
\end{tabular}
\hspace{3mm}
\begin{tabular}{|c|c|c|c|}
\hline
& \multicolumn{3}{c|}{{\bf\Red Flavor ratios $R_\nu$\Black}} \\
\hline
$\Delta E_\nu$ & CORT & CORT+TARGET & FLUKA 3D \\
   (GeV)       &      &             &          \\
\hline         &      &             &          \\
  0.1--0.2     & 0.52 &    0.51     &  0.48    \\
  0.2--0.3     & 0.52 &    0.50     &  0.49    \\
  0.3--0.5     & 0.51 &    0.50     &  0.50    \\
  0.5--0.7     & 0.50 &    0.50     &  0.50    \\
  0.7--1.0     & 0.49 &    0.50     &  0.50    \\
  1.0--2.0     & 0.47 &    0.49     &  0.49    \\
  2.0--3.0     & 0.44 &    0.47     &  0.45    \\
  3.0--5.0     & 0.40 &    0.43     &  0.42    \\
  5.0--7.0     & 0.36 &    0.37     &  0.38    \\
  7.0--10.0    & 0.32 &    0.32     &  0.34    \\
 10.0--20.0    & 0.26 &    0.26     &  0.29    \\
 20.0--30.0    & 0.20 &    0.19     &  0.23    \\
               &      &             &          \\
\hline
\end{tabular}
\end{table*}

\clearpage 

The AN angular distributions calculated with CORT and with FLUKA
3D code \citep{Battistoni00,Battistoni01b} are in good agreement
for any zenith angle at energies above 0.7--0.8 GeV. On the
other hand, CORT is systematically lower than FLUKA for
$E_\nu\lesssim1$~GeV. At the energy $E_\nu=0.5$~GeV the FLUKA
$4\pi$ averaged $\nu_e$, $\overline{\nu}_e$, $\nu_\mu$ and
$\overline{\nu}_\mu$ fluxes exceed the corresponding CORT results
by about 15\%, 27\%, 18\% and 18\%, respectively. This discrepancy
is only partially due to 3D effects (Fig. \ref{f:K-GS}), which can
account for an increase $\lesssim10\%$, and it is most probably
related to differences between the hadronic interaction models
adopted in FLUKA and CORT codes.

It is pertinent to note that several recent 3D Monte Carlo
calculations of the AN flux
\citep{Honda01a,Honda01b,Liu01,Plyaskin01,Tserkovnyak01,Wentz01a}
confirmed the sharp enhancement of low-energy neutrino intensities
for near-horizontal directions predicted with FLUKA. However
the quantitative characteristics of this enhancement are very
model-dependent and vary over a broad range from one calculation
to another.

\protect\subsection*{\em Prompt neutrinos}

\begin{figure*}[t]
\centering
\ifpdf
\includegraphics[width=0.88\linewidth]{e1t.pdf} 
\includegraphics[width=0.88\linewidth]{e2t.pdf} 
\else
\includegraphics[width=0.88\linewidth]{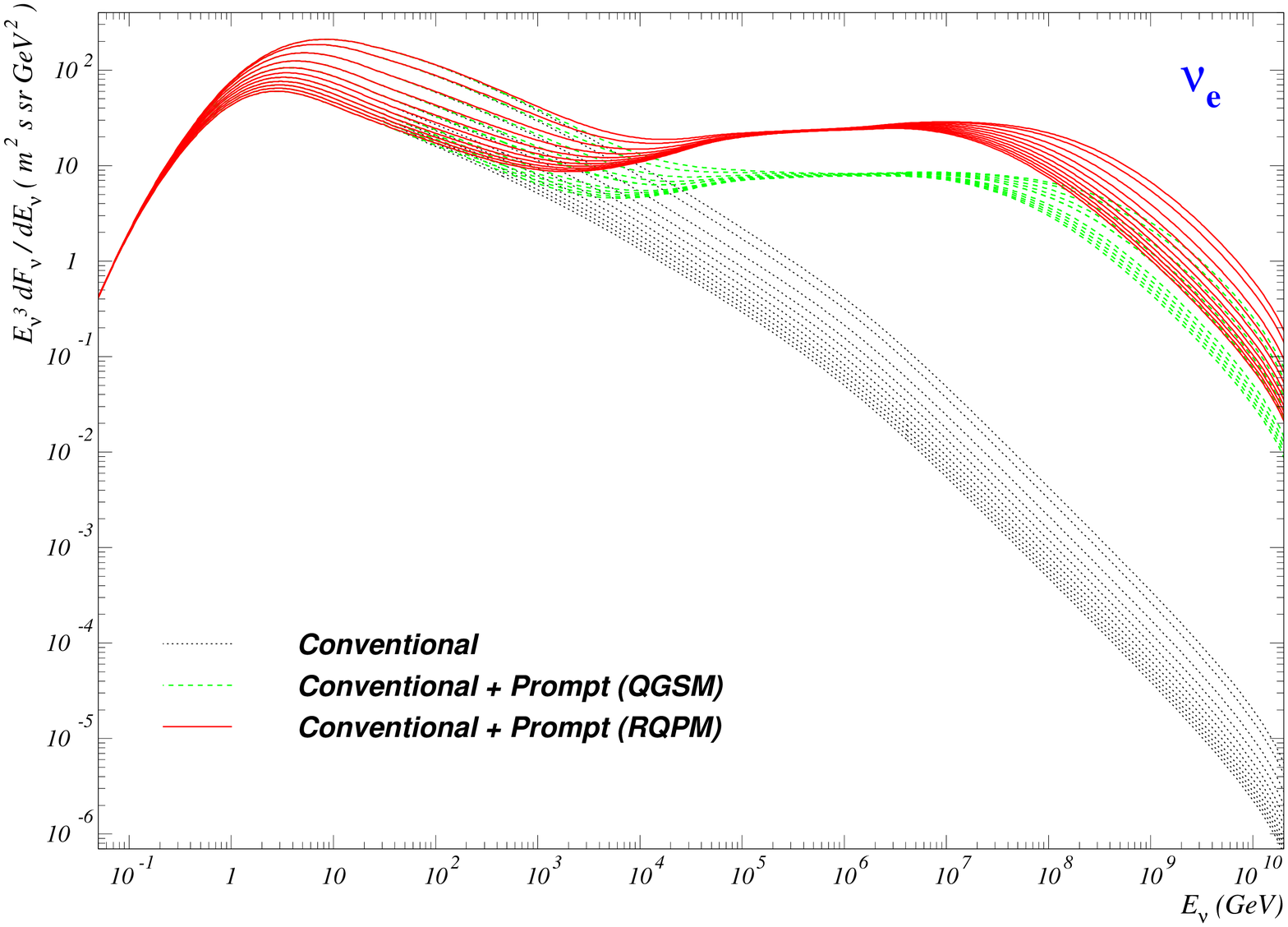} 
\includegraphics[width=0.88\linewidth]{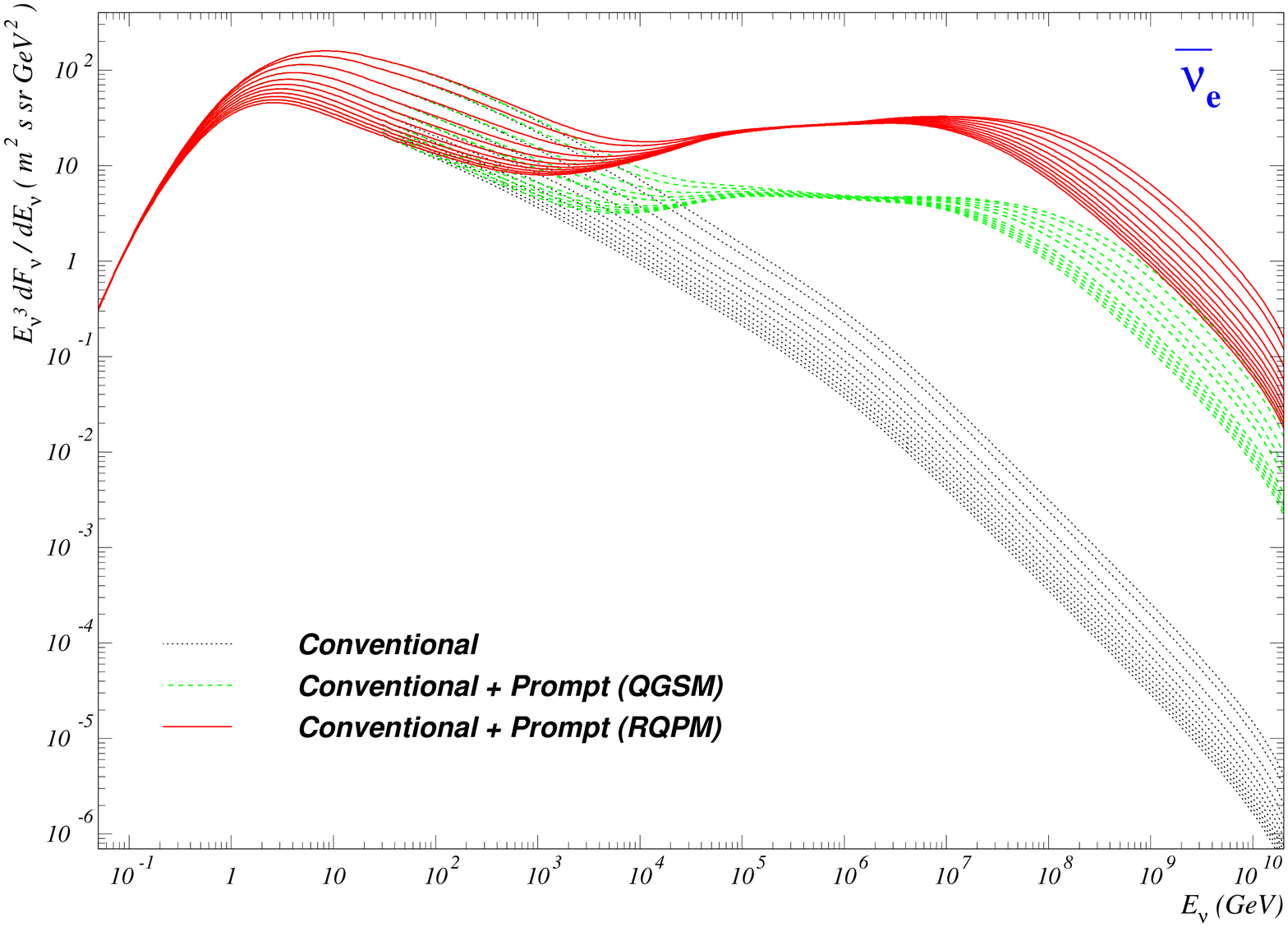} 
\fi
\protect\caption{Differential energy spectra of electron neutrinos
                 and antineutrinos for 11 zenith angles $\theta$.
                 Low-energy range is for Kamioka site. Contributions
                 from charm decay are calculated using recombination
                 quark-parton model (RQPM) and quark-gluon string
                 model (QGSM). At high energies, from smallest to
                 largest fluxes, $\cos\theta$ varies from 0 to 1
                 with an increment of 0.1 for each group of curves.
\label{f:Nu_e}}
\end{figure*}

\begin{figure*}[t]
\centering
\ifpdf
\includegraphics[width=0.88\linewidth]{m1t.pdf} 
\includegraphics[width=0.88\linewidth]{m2t.pdf} 
\else
\includegraphics[width=0.88\linewidth]{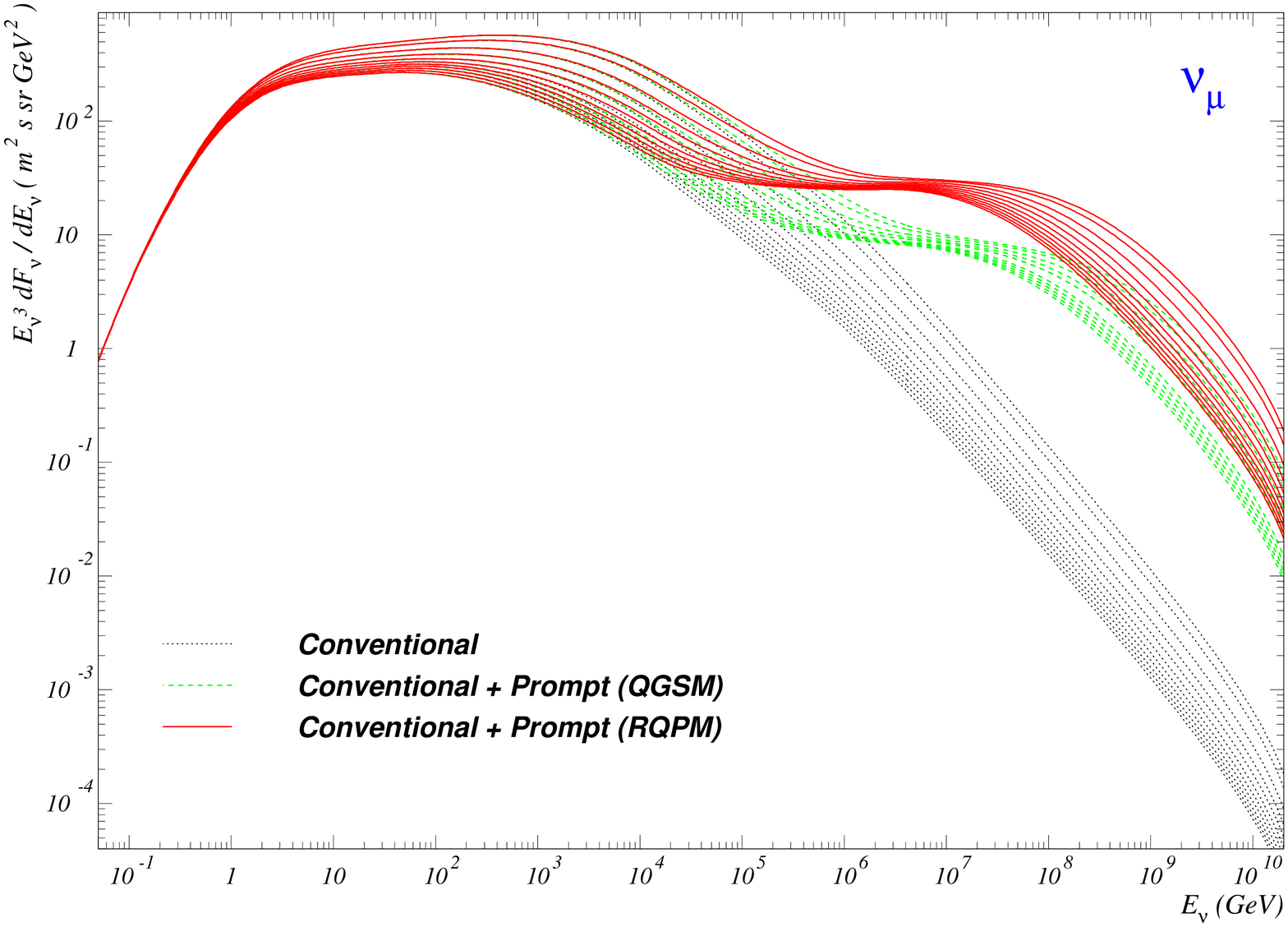} 
\includegraphics[width=0.88\linewidth]{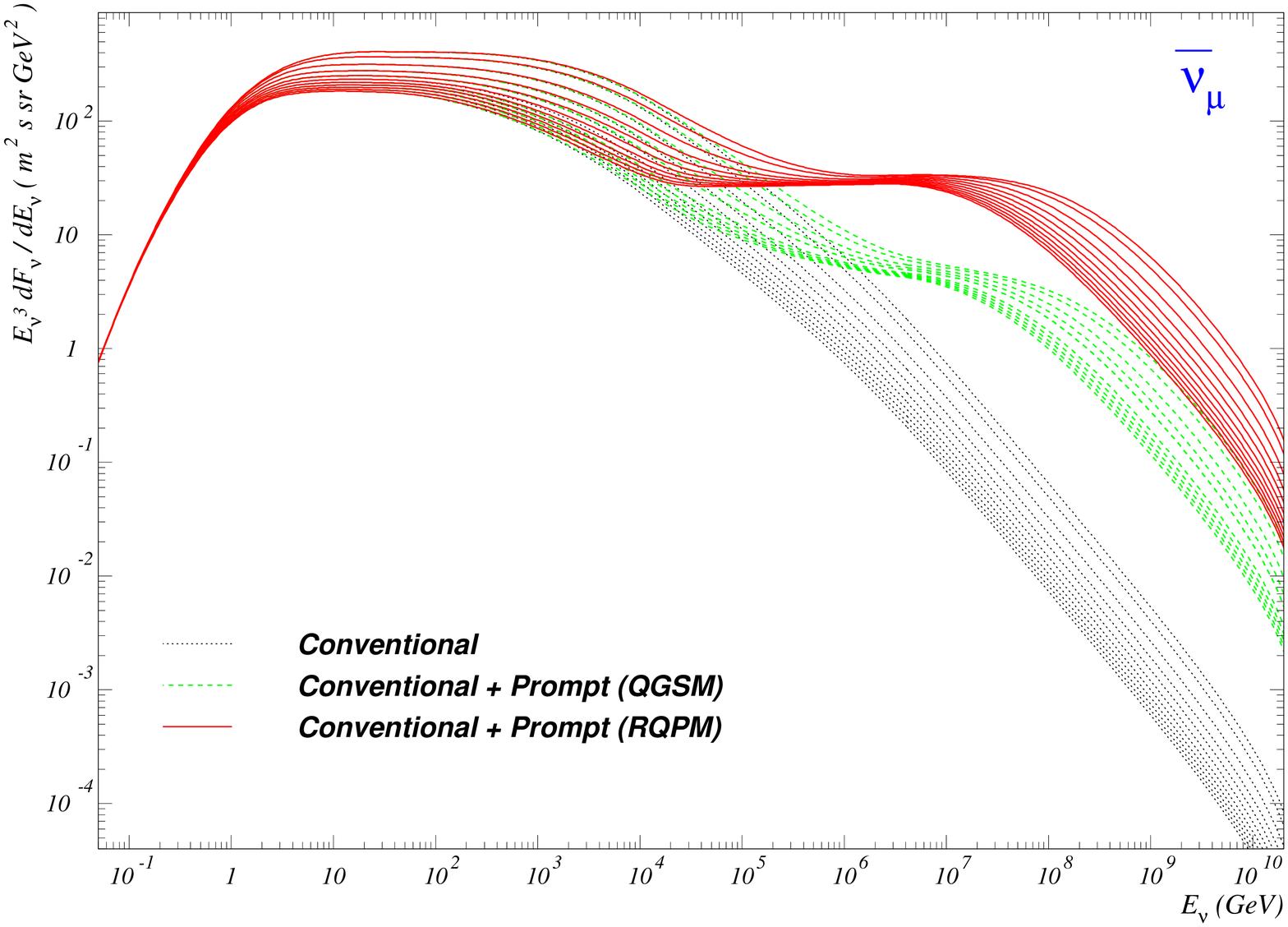} 
\fi
\protect\caption{Differential energy spectra of muon neutrinos
                 and antineutrinos for 11 zenith angles $\theta$.
                 Low-energy range is for Kamioka site. Contributions
                 from charm decay are calculated using recombination
                 quark-parton model (RQPM) and quark-gluon string
                 model (QGSM). At high energies, from smallest to
                 largest fluxes, $\cos\theta$ varies from 0 to 1
                 with an increment of 0.1 for each group of curves.
\label{f:Nu_m}}
\end{figure*}

Figures \ref{f:Nu_e} and \ref{f:Nu_m} collect the differential
energy spectra of downward going atmospheric neutrinos calculated
within a wide energy range (from 50 MeV to about 20 EeV) for 11
zenith angles. Figures show the ``conventional'' neutrino
contribution (originated from decay of pions, kaons and muons)
and the total AN spectra which include the ``prompt'' neutrino
contribution originated from semileptonic decays of charmed hadrons
(mainly $D^\pm$, $D^0$, $\overline{D}^0$ mesons and $\Lambda_c^+$
hyperons).%
\footnote{We do not discuss here the contributions from $D_s$ mesons
          and heavier states (such as $b\overline{b}$,
          $t\overline{t}$, $W$, $Z$) which become important sources
          of atmospheric (and extraterrestrial) prompt $\tau$ neutrinos
          at very high energies \citep{Pasquali99b,Athar01}.}
The prompt neutrino contribution must dominate at very high energies.
However the charm hadroproduction cross sections are very
model-dependent and cannot be unambiguously predicted for lack of a
generally accepted model. As a result the prompt neutrino
contribution and even the energies above which the prompt muon and
electron neutrinos become dominant are very uncertain as yet
\citep{Bugaev98,Pasquali98,Pasquali99a,Gelmini00a,Gelmini00b,%
Volkova01,Costa01a,Costa01b}.
The results shown in Figs. \ref{f:Nu_e} and \ref{f:Nu_m} are obtained
by using the two phenomenological approaches to the charm production
problem: the quark-gluon string model (QGSM) and recombination
quark-parton model (RQPM). The prompt muon fluxes predicted by QGSM
and RQPM are both consistent with the current deep underground data
and may be considered as the safe lower and upper limits for the
prompt muon contributions.
The basic assumptions of these models and the aspects of atmospheric
cascade calculations were described by \citet{Bugaev89b}.%
\footnote{See also Refs. \citep{Bugaev98,Naumov98a} for more details
          and related references.}

Calculations of conventional AN fluxes for energies below
$\sim100$ GeV are done with CORT for Kamioka site, while the
high-energy part of the conventional neutrino spectra as well as
the prompt neutrino contributions are obtained by using the results
of Ref. \citep{Naumov98b}. Needless to say the high-energy AN fluxes
are independent of location. The high-energy calculation takes into
account many ``thin'' effects, like $K_{e3}$ and $K_{\mu3}$ form
factors and $K_S^0$ semileptonic decays, meson regeneration and
charge exchange through reactions
$\pi^\pm+\mathrm{Air}\to\pi^{\pm(\mp)}+X$,
$\pi^\pm+\mathrm{Air}\to K^{\pm(0)}+X$, etc., as well as through
$K_{\pi2}$ and $K_{\pi3}$ decays.
The low- and high-energy parts of the spectra are smoothly merged
by using a polynomial (over $\log(E_\nu)$) clutching functions.
The results of these calculations for all underground detectors
are now embedded in a simple FORTRAN code ``CORTout''. It is based
on a two-dimensional spline interpolation over the detailed tables
and it may be useful for a fast evaluating the energy spectra and
zenith-angle distributions of conventional and prompt atmospheric
neutrinos within the energy range from 0.05 to $\sim10^{10}$ GeV.
The code is available upon request from the author.

\protect\section{Conclusions}
\label{sec:Conclusions}

Let us briefly re-state the main conclusions which follow from the
outcome of \citet{Fiorentini01a,Fiorentini01b} and from the present
study.

The results of extensive calculations performed with the new version
of code CORT by using the updated KM+SS hadronic interaction model
and the BESS+JACEE fit for the spectrum and composition of primary
cosmic rays are in substantial agreement with a quite representative
set of data on muon momentum spectra and charge ratios measured at
different atmospheric depths, zenith angles, and geomagnetic
locations.
This provides a solid evidence for the validity of our description
of hadronic interactions and shower development.
The low-energy atmospheric neutrino fluxes calculated with CORT for
many geomagnetic locations are systematically lower than those used
in the current analyses of the data from underground neutrino
experiments, while the neutrino flavor ratio and ``up-to-down''
asymmetry are essentially the same as in all recent calculations.
It is difficult to increase the AN flux without spoiling the
agreement with the current data on hadronic interactions, primary
spectrum and muon fluxes.


\begin{acknowledgements}
It is pleasure to thank Gianni Fiorentini and Francesco Villante
for their cooperation and for many constructive discussions. I thank
Kostya Kuzmin and Segey Sinegovsky for significant contribution.
I am thankful to Mirko Boezio, Patricia Hansen, Pierre Le Coultre,
and Tomonori Wada for providing me with the data from
CAPRICE\,94/98, L3+C, and Okayama experiments before publication.
I also thank Thomas Hebbeker and Charles Timmermans for their
kind assistance in finding and understanding many experimental
data. This research was supported in part by the Ministry of
Education of the Russian Federation, the program ``Universities of
Russia'' and by the Florence L3+C group.
\end{acknowledgements}

\clearpage 

\clearpage 


\end{document}